\documentclass[english,a4paper,a4paper,aps,prb,twocolumn]{revtex4}
\usepackage{ae}
\usepackage{aecompl}
\usepackage[T1]{fontenc}
\usepackage[latin1]{inputenc}
\usepackage{array}
\usepackage{verbatim}
\usepackage{amsmath}
\usepackage{graphicx}
\usepackage{amssymb}

\makeatletter

\providecommand{\tabularnewline}{\\}

\usepackage{array}

\makeatletter

\usepackage[usenames]{color}

\makeatother

\usepackage{babel}
\makeatother
\begin{document}

\title{Comparison of Dissipative Particle Dynamics and Langevin thermostats\\
 for out-of-equilibrium simulations of polymeric systems}

\author{C. Pastorino$^{\ddag}${*}, T. Kreer{*}{*}, M. Müller$^{\dag}$,
and K. Binder{*}}

\affiliation{$\ddag$Departamento de Física, Centro Atómico Constituyentes, CNEA/CONICET,
Av.Gral.~Paz 1499, 1650 Pcia.~de Buenos Aires, Argentina}

\affiliation{{*}Institut für Physik WA331, Johannes Gutenberg-Universität, 55099,
Mainz, Germany}

\affiliation{{*}{*}Institut Charles Sadron, 6 Rue Boussingault, 67083 Strasbourg,
France}

\affiliation{$^{\dag}$ Institut für Theoretische Physik, Friedrich-Hund-Platz
1, 37077 Göttingen, Germany}

\begin{abstract}
In this work we compare and characterize the behavior of Langevin
and Dissipative Particle Dynamics (DPD) thermostats in a broad range
of non-equilibrium simulations of polymeric systems. Polymer brushes
in relative sliding motion, polymeric liquids in Poiseuille and Couette
flows, and brush-melt interfaces are used as model systems to analyze
the efficiency and limitations of different Langevin and DPD thermostat
implementations. Widely used coarse-grained bead-spring models under
good and poor solvent conditions are employed to assess the effects
of the thermostats. We considered equilibrium, transient, and steady
state examples for testing the ability of the thermostats to maintain
constant temperature and to reproduce the underlying physical phenomena
in non-equilibrium situations. The common practice of switching-off
the Langevin thermostat in the flow direction is also critically revisited.
The efficiency of different weight functions for the DPD thermostat
is quantitatively analyzed as a function of the solvent quality and
the non-equilibrium situation. 
\end{abstract}
\maketitle

\section{Introduction}

The need and use of thermostats in computer simulations started with
the beginning of the field itself. The original Molecular Dynamics
(MD) method, intended for microcanonical ensemble simulations, was
soon extended to different ensembles in order to mimic conditions
in which experiments are actually performed. The thermostat in MD
simulations implies the assumption that the system transports heat
{}``instantaneously'' fast on the spatial scale of the simulation.
Even when this is arguably not completely correct in a real system
many studies are faced with the situation of performing simulations
at constant temperature as a way of obtaining a physically meaningful
condition.

It is a non-trivial challenge to achieve a constant temperature in
a simulation of driven soft matter systems, such as polymer brushes
interacting with flowing polymer melts, which we consider in the present
work.

In this work, we present simulation data which we obtained by integrating
Langevin-like equations with the standard MD integrators using either
a Langevin or a DPD thermostat to maintain temperature. \textbf{}This
is often called \textbf{}\emph{Stochastic Dynamics} (SD) or \textbf{}\emph{Brownian
Dynamics} (BD)\cite{tildesley,bd_sd,huenenberger_rev_thermostats},
but we address a physical regime in which the friction and stochastic
forces, added to the conservative forces of the system, are small
enough in order not to significantly perturb the natural dynamics
of the polymeric system. This is typically achieved by using the smallest
possible value for the friction constant, $\gamma$, providing that
the temperature is conserved under the desired physical conditions.
This approach also requires that the system has intermediate to high
monomer densities to warrant that the friction due to the conservative
interactions is significantly larger than the frictional and random
forces arising from the thermostat. The regime of dilute polymer solutions
is excluded from this study because, in this case, the polymer-solvent
interactions are important for the local dynamics of the molecules. 

In this sense we consider the Langevin and DPD frictional and stochastic
forces as \textbf{}\emph{thermostats} that are added to the true conservative
forces of the system, as it is generally employed in MD. This physical
regime is of great interest for a number of soft matter systems, such
as polymeric interfaces, blends and melts, that are successfully studied
in the framework of coarse-grained models\cite{grest_review_brush,priezjev1,priezjev2,dpd3}.

%
{}{}

A case which has attracted abiding interest is the simulation of out-of-equilibrium
phenomena in which a rate of energy must be injected into the system
to drive it out of equilibrium. This energy must be removed in order
to keep the temperature constant, and this is usually done by the
action of a thermostat. There are excellent reviews that describe
the different types of thermostats and their respective advantages
and limitations for studying various systems and physical phenomena.
\cite{tildesley,understanding,huenenberger_rev_thermostats} Of course,
it is crucial that the dynamical behavior observed in a simulation
faithfully represents the actual dynamics of the desired system and
is essentially free from artifacts introduced by the thermostat. This
issue shall be explored for out-of-equilibrium simulations of polymeric
systems in our study.

In this article, we focus on the behavior of Langevin and DPD\cite{dpd1,dpd2,dpd4,dpd5}
thermostats for a range of typical polymeric systems in non-equilibrium
conditions. The former has been widely used in equilibrium simulations
but is known to have undesirable properties, such as screening of
hydrodynamic interactions and lack of Galilean invariance,\cite{duenweg2,dpd3}
in non-equilibrium situations. A typical workaround when using the
Langevin thermostat for non-equilibrium simulations consists in switching-off
the thermostat in the direction in which non-conservative external
forces are applied to the system or applying it only in one spatial 
direction. In this way one recovers momentum conservation in the
shear direction, while conserving the temperature by applying the
thermostat in the perpendicular direction in which no direct non-conservative
force is applied.\cite{thompson_robbins,priezjev2,priezjev1}

The DPD scheme only recently has started to be utilized as a standalone
thermostat.\cite{dpd3,claudio1} It was originally developed as a
method for performing meso-scale simulations by combining this thermostat
and very {}``soft'' potentials. The latter allow for the use of
a large time step in MD simulations.\cite{dpd1,dpd2,dpd4} The maximal
time step that is permissible in DPD simulations has been investigated
thoroughly. \cite{understanding,dpd_vattulainen,dpd_bessold_vattulainen,dpd_pagonabarraga_frenkel}
Utilizing a DPD thermostat in conjunction with {}``hard'' potentials
-- typical of coarse-grained models widely used for polymers and other
condensed matter systems\cite{kremer_grest} -- one looses this advantage,
and one must take a time step on the order of that typically used
in MD simulations of coarse-grained models. The local conservation
of momentum and the Galilean invariance, however, are inherited from
the original DPD method, and possibly this is a great advantage.

In the following, we consider three polymeric systems to assess the
effects of two versions of Langevin thermostats (with and without
switching-off the thermostat in the flow direction) and the DPD thermostat:
(a) single end-grafted polymer layers (so-called polymer brushes),
(b) two opposing and interdigitating polymer brushes, and (c) a brush-melt
interface, which exhibits a rich wetting behavior. The equilibrium
properties of these reference systems are interesting and have been
comprehensively studied in previous works\cite{grest_review_brush,droplets_marcus_luis2,Kreer2001,Kreer2003,Kreer2004,luis06,claudio1}.
The brush layers are characterized by the number of grafted chains
per unit area or grafting density $\rho_{g}$. Typical equilibrium
density profiles for these three systems are shown in Fig.~\ref{fig:Equilibrium-density-profiles}.
Additionally, a simple bulk system of a polymeric liquid with periodic
boundary conditions in all three directions was considered.

These systems are also of great interest out of equilibrium, for instance,
because of the surprisingly small friction of two opposing brushes
sliding past each others.\cite{jacob_klein,jacob_klein2,Kreer2001}

The interface between a brush and a melt of identical chains is a
prototypical example for a copolymer-laden interface or a melt in
contact with a soft, elastically deformable substrate (e.g., confining
brush-coated walls of a channel).\cite{claudio1} In addition to the
rich wetting properties, typical applications (e.g., droplet break-up
in a polymer blend under shear or flow in a microfluidic channel)
involve flow and shear at the brush-melt interface. Therefore, the
study of boundary conditions and the rheological properties of the
macromolecular liquid subjected to different types of flows make the
non-equilibrium properties of this system particularly interesting.

The behavior of the thermostats in equilibrium, transient, and different
kinds of steady states was tested in a wide number of typical situations
that can be encountered in simulations. Also, Poiseuille and Couette
flows of the polymeric liquid were considered to compare different
weight functions of the DPD thermostat.

The details of the thermostats and the polymer model are explained
in section \ref{sec:model}. Section \ref{sec:Results} presents the
discussion of our results, which begins with a quantitative study
of the relative strengths of thermostats for a given set of parameters.
This section is divided in subsections corresponding to each different
system: the analysis of single-brush layer transient states is presented
in section \ref{sub:Single-brush}, and the steady state of two polymer
brushes in relative sliding motion is discussed in \ref{sub:brush-brush}
for two different grafting densities. In this way, we address two
regimes: concentrated solution or melt in which hydrodynamic interactions
are screened and the a more dilute regime. Finally, the study of different
DPD weight functions and their efficiency for conserving the temperature
in the strong out-of-equilibrium regime for Couette and Poiseuille
flows of the brush-melt interface is described in subsection \ref{sub:role_weight_functions}.
Discussion and concluding remarks are presented in section \ref{sec:Conclusions}.

\section{Polymer Model and thermostat details\label{sec:model}}

We used a well established coarse-grained bead-spring model\cite{kremer_grest}
for polymers with excluded volume and intra-molecular interactions.
This model has been applied to a variety of thermodynamic conditions,
chain lengths, and physical regimes such as glasses, melts, dilute
solutions, etc.\cite{review_kroeger,grest_review_brush,review_baschnagel_varnik,duenweg_kremer_solution}
The interaction between neighboring beads along the same polymer is
modeled by a finite extensible non-linear elastic (FENE) potential:
\begin{equation}
U_{{\rm FENE}}=\begin{cases}
-\frac{1}{2}k\,\, R_{0}^{2}\ln\left[1-\left(\frac{r}{R_{0}}\right)^{2}\right] & r\leq R_{0}\\
\infty & r>R_{0}\end{cases},\end{equation}
 where the maximum allowed bond length is $R_{0}=1.5\sigma$, the
spring constant is $k=30\varepsilon/\sigma^{2}$, and $r=|{\mathbf{r}_{i}}-{\mathbf{r}_{j}}|$
denotes the distance between neighboring monomers. Excluded volume
interactions at short distances and van-der-Waals attractions between
segments are described by a truncated and shifted Lennard-Jones (LJ)
potential: \begin{equation}
U(r)=U_{{\rm LJ}}(r)-U_{{\rm LJ}}(r_{{\rm c}})\,,\end{equation}

with

\begin{equation}
U_{{\rm LJ}}(r)=4\varepsilon\left[\left(\frac{\sigma}{r}\right)^{12}-\left(\frac{\sigma}{r}\right)^{6}\right]\,,\end{equation}
 where the LJ parameters, $\varepsilon=1$ and $\sigma=1$, define
the units of energy and length, respectively. The temperature is therefore
given in units of $\varepsilon/k_{B}$, with $k_{B}$ the Boltzmann
constant. $U_{{\rm LJ}}(r_{{\rm c}})$ is the LJ potential evaluated
at the cut-off radius. We considered two values as cut-off distance:
(i) twice the minimum of the LJ potential: $r_{{\rm c}}=2\times2^{\frac{1}{6}}\simeq2.24\sigma$,
which allows to consider poor solvent conditions, and (ii) $r_{{\rm c}}=2^{\frac{1}{6}}\simeq1.12\sigma$,
which models good solvent conditions. In the latter case, the interactions
between monomers of different chains are purely repulsive,\cite{Kreer2001,Kreer2003,Kreer2004}
whereas in the former case, longer ranged attractions are included
giving rise to liquid-vapor phase separation and droplet formation\cite{MacDowell2000,droplets_marcus_luis1}
below the $\Theta$-temperature, $\Theta=3.3\varepsilon/k_{B}$. We
analyze the efficiency of the thermostats for both cases.

The substrate is modeled as an idealized flat and impenetrable wall,
which interacts with the polymer segments via an integrated Lennard-Jones
potential :

\begin{equation}
V_{{\rm wall}}(z)=|A|\left(\frac{\sigma}{z}\right)^{9}-A\left(\frac{\sigma}{z}\right)^{3},\label{eq:v_wall}\end{equation}
 where $A=3.2\varepsilon$, used throughout this work, is sufficient
to make the liquid wet the bare substrate.\cite{droplets_marcus_luis1,Muller2003}
The tethered beads are fixed randomly in the grafting plane at a distance
of $1.2\sigma$ from the wall position for all the cases. In the following,
we will use LJ units\cite{tildesley} for all quantities, unless explicitly
mentioned otherwise.

\subsection{Langevin and Dissipative Particle Dynamics thermostats}

Both, Langevin and DPD thermostats, can be written in a general form,
starting from the Hamiltonian equations of motion:\cite{understanding,tildesley}
\begin{eqnarray}
\dot{\mathbf{r}_{i}} & = & \frac{\mathbf{p}_{i}}{m_{i}}\nonumber \\
\dot{\mathbf{p}_{i}} & = & \mathbf{F}_{i}+{\mathbf{F}_{i}}^{{\rm D}}+{\mathbf{F}_{i}}^{{\rm R}},\label{eq:md_eq_motion}\end{eqnarray}
 where $\mathbf{F}_{i}$ is the total conservative force on each particle
and $m_{i}$ the mass of each particle. ${\mathbf{F}_{i}}^{{\rm D}}$
and ${\mathbf{F}_{i}}^{{\rm R}}$ are the forces due to the thermostat
and will be of different form for Langevin and DPD cases. The difference
between DPD and Langevin thermostats is the way in which random and
dissipative forces are applied.

In the case of Langevin thermostats, the dissipative force on particle
$i$ is given by ${\mathbf{F}_{i}}^{{\rm D}}=-\gamma\mathbf{v}_{i}$,
where $\gamma$ is the friction coefficient and $\mathbf{v}_{i}$
the particle velocity. The random force, $\mathbf{F}_{i}^{{\rm R}}$,
has zero mean value and its variance satisfies\cite{huenenberger_rev_thermostats}

\begin{equation}
\langle F_{i\mu}^{{\rm R}}(t)F_{j\nu}^{{\rm R}}(t^{\prime})\rangle=2\gamma Tk_{B}\delta_{ij}\delta_{\mu\nu}\delta(t-t^{\prime})\,,\label{eq:mean-square-random-lgv}\end{equation}
 where the indices $i$ and $j$ label particles, $\mu$ and $\nu$
Cartesian components, and $T$ is the temperature at which the system
is simulated.

For the DPD case,\cite{dpd1,dpd2} the dissipative and frictional
forces are applied in a pair-wise form, such that the sum of thermostating
forces acting on a particle pair equals zero. The expression for the
forces is the following: \begin{eqnarray}
{\mathbf{F}_{i}}^{{\rm D}} & = & \sum_{j(\neq i)}{\mathbf{F}_{ij}^{{\rm D}}}\,\,;\,\,{\mathbf{F}_{ij}^{{\rm D}}}=-\gamma\omega^{{\rm D}}(r_{ij})(\hat{\mathbf{r}}_{ij}\cdot{\mathbf{v}_{ij}})\hat{\mathbf{r}}_{ij}\nonumber \\
{\mathbf{F}_{i}}^{{\rm R}} & = & \sum_{j(\neq i)}{\mathbf{F}_{ij}^{{\rm R}}}\,\,;\,\,{\mathbf{F}_{ij}^{{\rm R}}}=\sigma\omega^{{\rm R}}(r_{ij})\theta_{ij}\hat{\mathbf{r}}_{ij},\end{eqnarray}
 where for each vector ${\mathbf{a}}$ we define ${\mathbf{a}_{ij}}\equiv{\mathbf{a}_{i}}-{\mathbf{a}_{j}}$,
$\gamma$ is the friction constant, and $\sigma$ the noise strength.
Friction and noise, $\gamma$ and $\sigma$, obey the relation $\sigma^{2}=2k_{B}T\gamma$,
and the associated weight functions satisfy the fluctuation-dissipation
theorem if the following relation is fulfilled:\cite{dpd4} \begin{equation}
[\omega^{{\rm R}}]^{2}=\omega^{{\rm D}}.\label{eq:dpd_weights_rel}\end{equation}
 $\theta_{ij}$ is a random variable with zero mean and second moment
\begin{equation}
\langle\theta_{ij}(t)\theta_{kl}(t^{'})\rangle=(\delta_{ij}\delta_{jl}+\delta_{il}\delta_{jk})\delta(t-t^{'}).\end{equation}
 The standard weight functions found in the literature are: \begin{equation}
[\omega^{{\rm R}}]^{2}=\omega^{{\rm D}}=\begin{cases}
(1-r/r_{{\rm c}})^{2},r<r_{{\rm c}} & ,\\
0,\,\,\,\, r\geq r_{{\rm c}}\end{cases}\label{eq:usual-weight}\end{equation}
 where $r_{{\rm c}}$ is the cut-off radius for a given molecular
model. However, we emphasize that Eq.~(\ref{eq:usual-weight}) is
just the typical choice when the DPD thermostat is employed in conjunction
with {}``soft'' potentials. For arbitrary models, one can choose
a different set of functions providing that they fulfill Eq.~(\ref{eq:dpd_weights_rel}),
and one can exploit this freedom to optimize the efficiency of the
thermostat for {}``hard'' potentials. In this work, we will use
the standard weights, but also test other possibilities, whose forms
are given in the first row of table \ref{tab:Ntp}. The equations
of motion {[}Eq.~(\ref{eq:md_eq_motion})] were integrated using
the velocity Verlet algorithm\cite{tildesley,understanding} with
a time step of $dt=0.002\tau$, where $\tau=\sigma(m/\varepsilon)^{1/2}$
denotes the time unit in terms of LJ parameters.

\begin{figure}[!h]
\begin{centering}
\begin{tabular}{c}
\includegraphics[clip,width=0.32\textwidth]{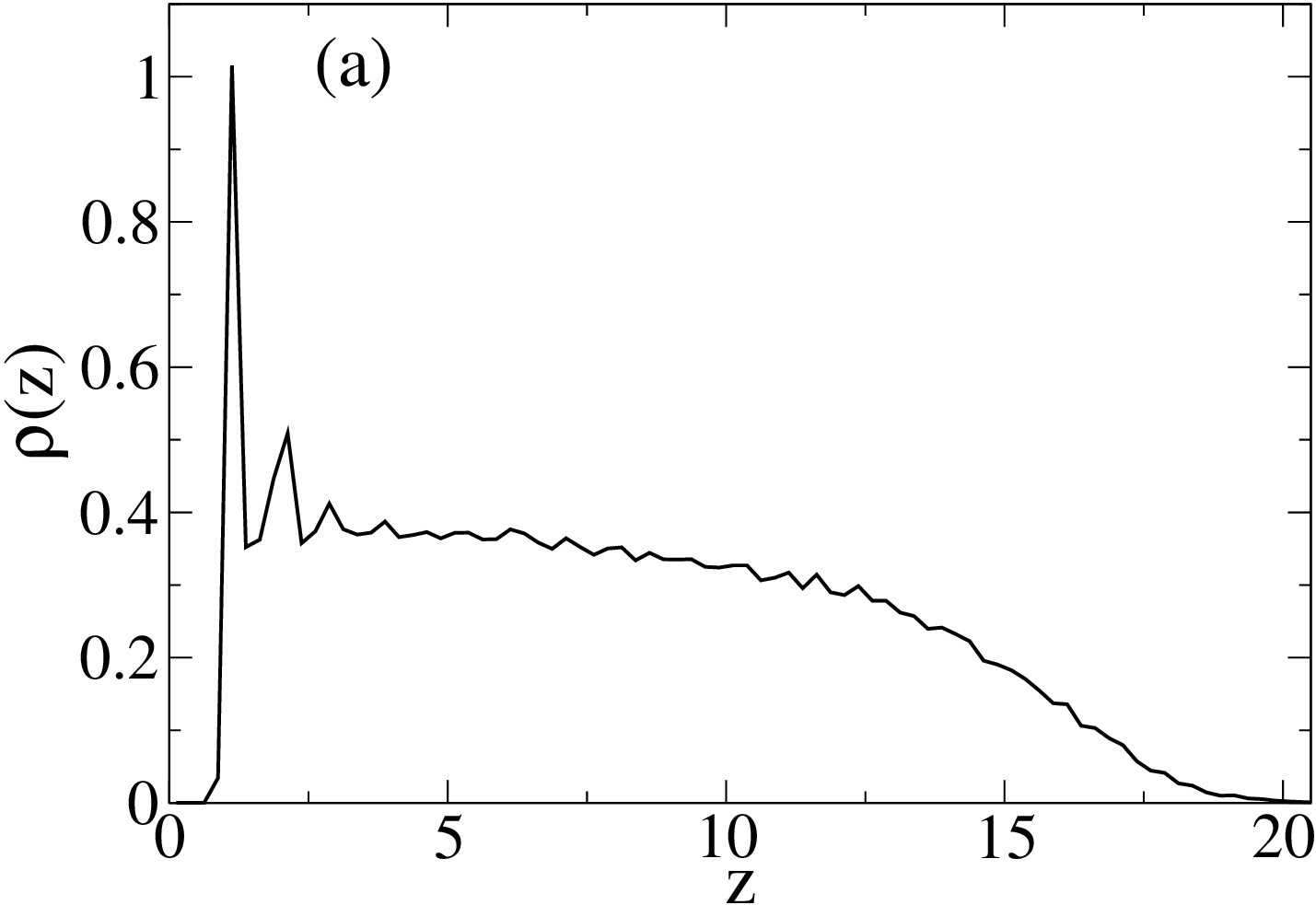}\tabularnewline
\includegraphics[clip,width=0.32\textwidth]{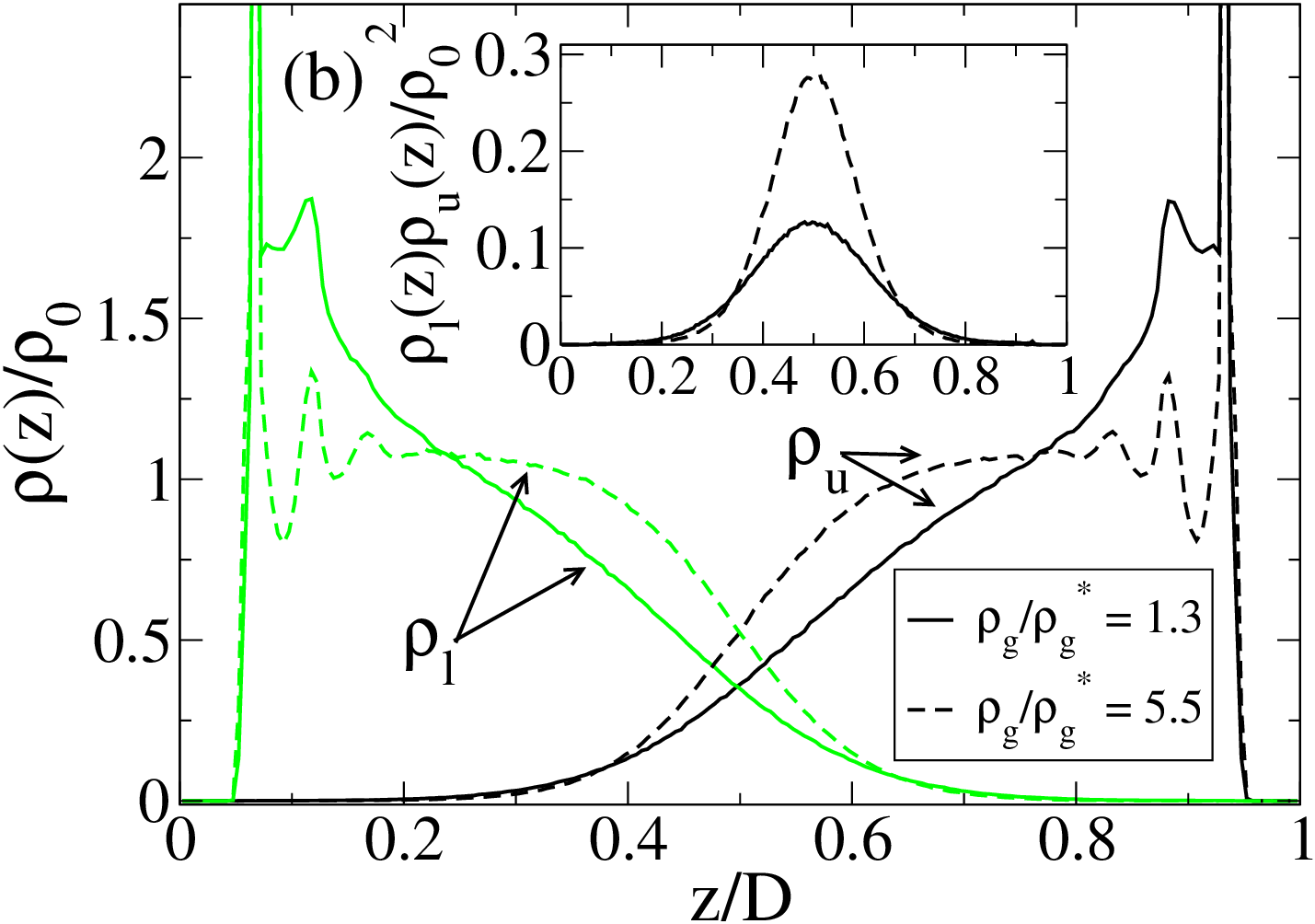}\tabularnewline
\includegraphics[clip,width=0.32\textwidth]{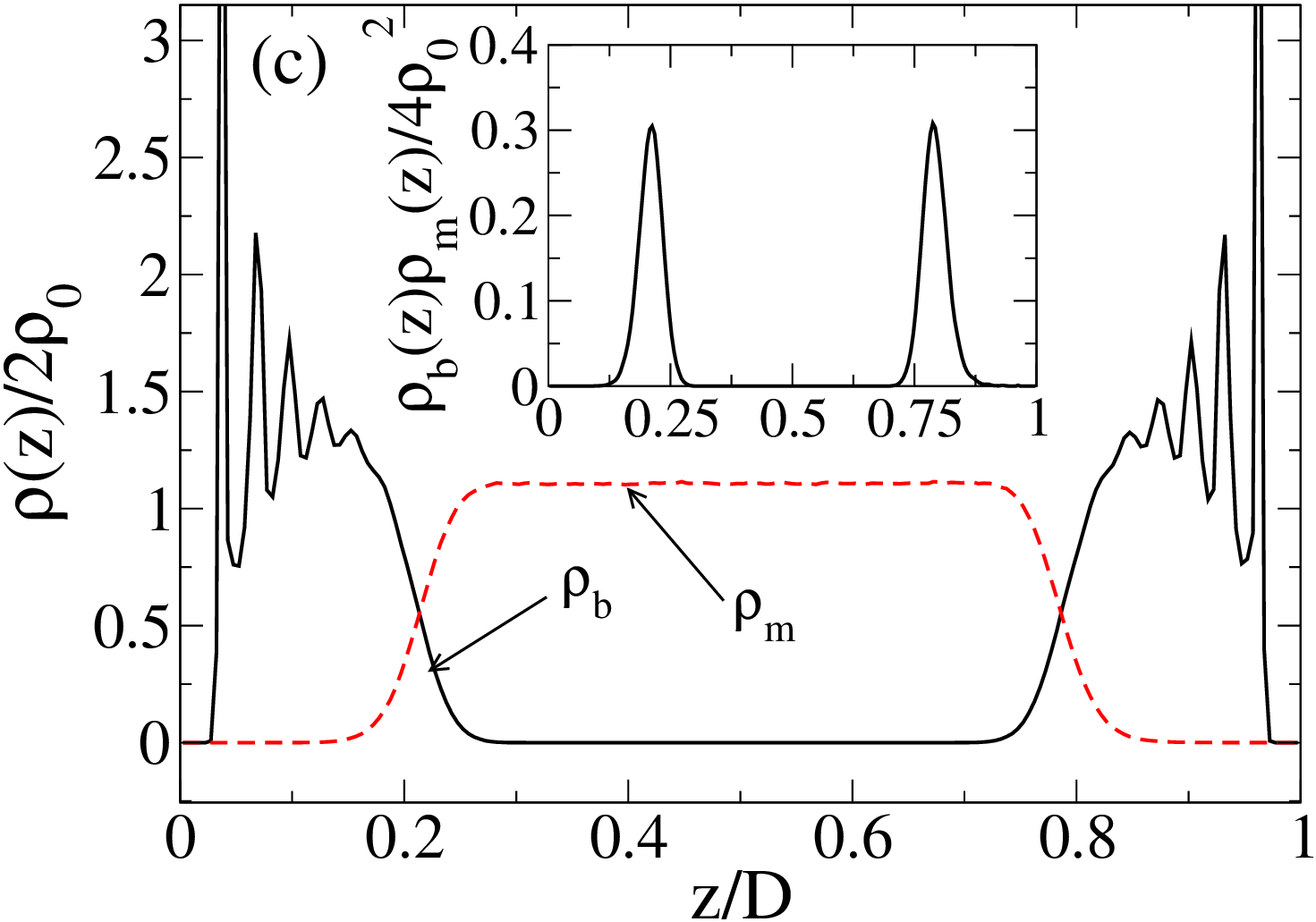}\tabularnewline
\end{tabular}
\par\end{centering}

\caption{(Color online) Equilibrium density profiles for the three systems
under consideration: (a) single brush layer, (b) two opposing brush
layers in interaction at two different grafting densities, and (c)
brush-melt interfaces. The insets show the product of upper and lower
brush density profiles $\rho_{{\rm u}}\times\rho_{{\rm l}}$ (b) or
the product of brush and melt phases $\rho_{b}\times\rho_{{\rm m}}$
(c), accounting for the level of interdigitation of the different
phases. In all cases, the temperature is $T=1.68$. The chain length
is $N=30$ for cases (a) and (b) and $N=10$ for case (c). The distance
between the grafted beads of the opposing brushes is $D=17.5$ for
case (b) and $D=30$ for case (c). The normalization is given by $\rho_{0}=\rho_{{\rm g}}/D$.}

\label{fig:Equilibrium-density-profiles} 
\end{figure}

\section{Results\label{sec:Results}}

\begin{figure}[t]
\begin{centering}
\includegraphics[clip,width=9cm]{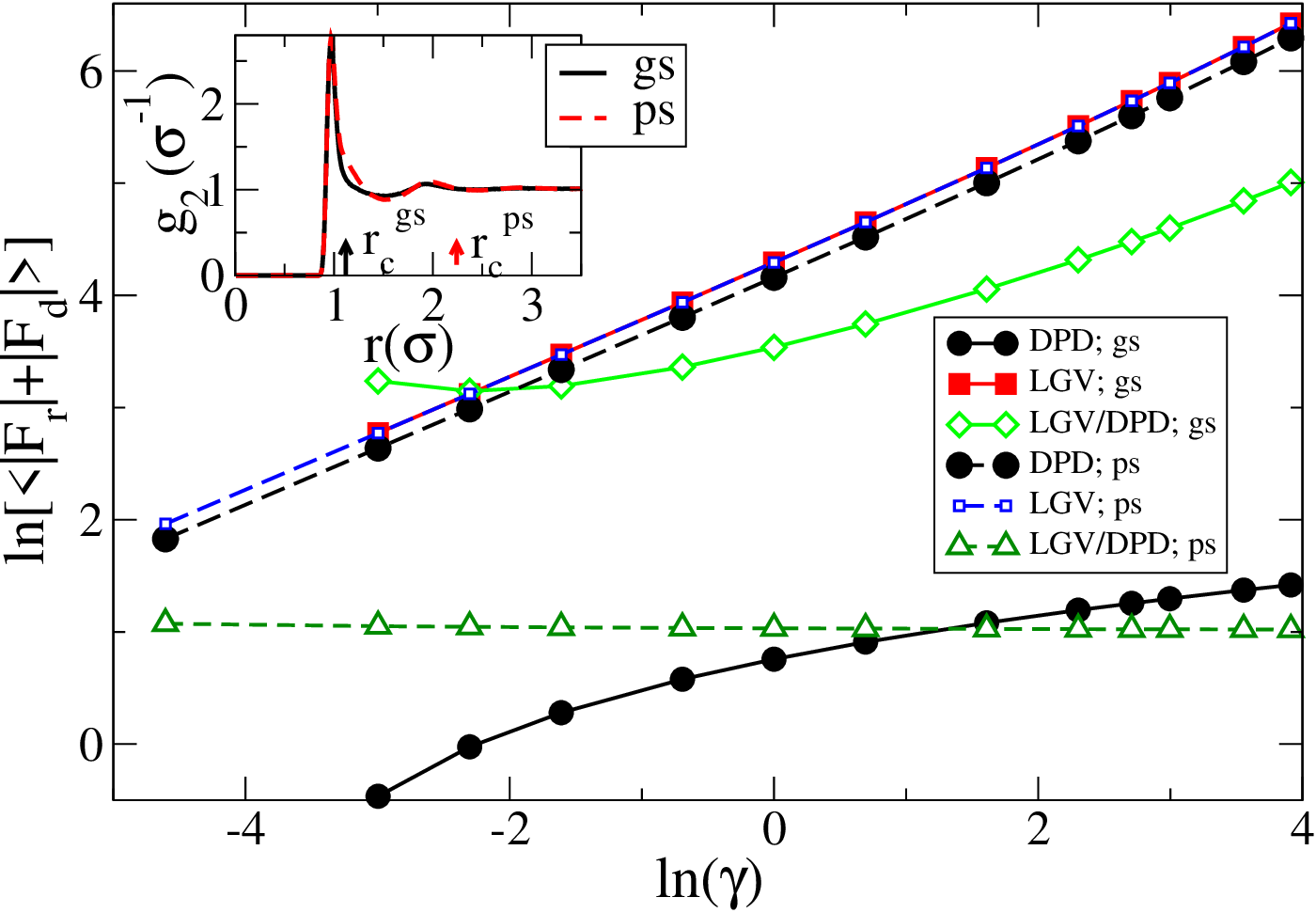} 
\par\end{centering}

\caption{(Color online) Comparison of mean random and dissipative forces for
DPD and Langevin (LGV) thermostats as a function of the friction constant,
$\gamma$, for a bulk system with temperature $T=1.68$ and density
$\rho=0.61$. The ratio $\frac{\langle F_{{\rm LGV}}\rangle}{\langle F_{{\rm DPD}}\rangle}$
is shown in open circles for both, good solvent (gs) and poor solvent
(ps) conditions. Inset: Pair correlation function $g_{2}(r)$ of the
polymeric liquid for good (solid line) and poor (dashed line) solvent
conditions. The arrows indicate the position of the cut-off radius
in each case. \label{fig:Comparison-of-gammas} }
\end{figure}

In this section, the results corresponding to the three polymeric
systems, whose equilibrium density profiles are shown in Fig.~\ref{fig:Equilibrium-density-profiles},
will be analyzed. The difference between weight functions and the
way random and friction forces are applied in Langevin and DPD thermostats
does not allow for a direct comparison of the friction strength $\gamma$
between both schemes. To obtain a direct measurement of thermostat
strengths, we computed the mean friction and dissipative forces as
a function of $\gamma$ for a polymeric liquid of 10-bead chains in
a bulk solution (using periodic boundary conditions in all spatial
directions). In the case of the DPD thermostat, the standard weight
functions were used {[}see Eq.~(\ref{eq:usual-weight})]. Figure
\ref{fig:Comparison-of-gammas} shows the total force as a function
of $\gamma$. Poor and good solvent conditions exhibit quite a different
behavior. In the first case, the Langevin and DPD thermostat show
a very similar behavior for the whole range of $\gamma$. For good
solvent conditions (only the repulsive part of the LJ potential is
kept), however, the mean Langevin forces are two orders of magnitude
larger than those of the corresponding DPD counterpart (for the standard
choice of weight functions). This means, for example, that a friction
constant $\gamma_{{\rm DPD}}=2$ is equivalent to a value of $\gamma_{{\rm LGV}}=0.01$,
as regards the mean value of the {}``thermostat force'' acting on
each bead. The ratio $\frac{\langle F_{{\rm LGV}}\rangle}{\langle F_{{\rm DPD}}\rangle}$
as a function of $\gamma$ is shown with open symbols for sake of
comparison. The reason for the big difference, in the case of good
solvent, is the structure of the liquid and, more important, the small
cut-off radius -- there are very few beads in the range of the weight
functions, and the standard weight functions are small in the vicinity
of $r_{{\rm c}}$ where most of the neighbors are located. The pair
correlation functions and corresponding cut-off radii are shown in
the inset of Fig.~\ref{fig:Comparison-of-gammas}.

At this point, it is important to recall that the original reason
for choosing those weight functions {[}Eq.~(\ref{eq:usual-weight})]
was based on the idea of using DPD together with {}``soft'' potentials\cite{dpd1,dpd5}
to achieve the largest possible time step. To this end, the thermostat
forces need also to be smoothly varying functions of the position
in order to have the same properties as the conservative forces. Actually,
the only constraint the weight functions must fulfill is the fluctuation-dissipation
theorem,\cite{dpd1,dpd4} i.e., Eq.~(\ref{eq:dpd_weights_rel}).

We will see below that the standard choice can be even bad for non-equilibrium
simulations, in which a significant amount of heat per unit time has
to be removed.

To quantify the efficiency of the thermostat to maintain constant
temperature, we define the number of thermostated particles, $N_{{\rm TP}}$,
for a given DPD weight function pair as \begin{equation}
N_{{\rm TP}}=\rho_{0}\int_{0}^{r_{{\rm c}}}\omega^{{\rm R}}(r)g_{2}(r)4\pi r^{2}{\rm d}r,\label{eq:Ntp}\end{equation}
 where $r_{{\rm c}}$ is the cut-off radius of the conservative potentials
which coincides with the cut-off of the weight functions for the dissipative
and frictional forces. $g_{2}(r)$ is the pair correlation function
for particles in the polymeric liquid. For the standard weight functions
and good solvent conditions, $N_{{\rm TP}}$ is rather small (cf.~Fig.~\ref{fig:Comparison-of-gammas}).

The pair correlation functions for poor and good solvent conditions
were taken from bulk simulations of a 10-bead polymeric liquid at
$T=1.68$ and $\rho=0.61$ which corresponds to the density of a melt
that coexists with its vapor.\cite{droplets_marcus_luis1} In table
\ref{tab:Ntp} the calculated values of $N_{{\rm TP}}$, as given
by Eq.~(\ref{eq:Ntp}), are shown for three different choices of
weight functions. The second column shows $N_{{\rm TP}}$ for the
standard weight functions {[}Eq.~(\ref{eq:usual-weight})]. The third
column shows the square root of the usual weight functions which slightly
increases the force in the region $r\lesssim r_{{\rm c}}$. The last
column corresponds to constant weight functions, i.e., $\omega^{{\rm R}}=\omega^{{\rm D}}=\Theta(r_{{\rm c}}-r)$,
with $\Theta$ being Heaviside's step function. The differences among
the weight functions is evident from the values of $N_{{\rm TP}}$.
Different weight functions give rise to significant changes in the
efficiency of the thermostat. In particular, the standard weight functions
inherited from DPD models with {}``soft'' potentials present the
lowest value for $N_{{\rm TP}}$ which is significantly smaller than
the values for other choices. As will be shown in section \ref{sub:role_weight_functions},
this is indeed an important issue in out-of-equilibrium simulations.

\begin{table}
\begin{centering}
\begin{tabular}{|>{\centering}p{2cm}|c|c|c|}
\hline 
$N_{{\rm TP}}(\omega^{{\rm R}})$&
$\omega^{{\rm R}}=1-\frac{r}{r_{{\rm c}}}$&
$\omega^{{\rm R}}=\sqrt{1-\frac{r}{r_{{\rm c}}}}$&
$\omega^{{\rm R}}=\Theta(r_{{\rm c}}-r)$\tabularnewline
\hline
\hline 
good solvent ($r_{{\rm c}}=1.12$)&
0.301&
0.904&
2.997\tabularnewline
\hline 
poor solvent ($r_{{\rm c}}=2.24$)&
6.753&
12.814&
28.868\tabularnewline
\hline
\end{tabular}
\par\end{centering}

\caption{Number of thermostated particles, $N_{{\rm TP}}$, for different
weight functions using the DPD thermostat. See text and Eq.~(\ref{eq:Ntp}). }

\label{tab:Ntp} 
\end{table}

\subsection{Single brush\label{sub:Single-brush}}

We consider a single polymer brush layer with $N=30$ beads per chain
in equilibrium under good solvent conditions. The brush stretches
freely according to the balance of entropy and steric repulsion between
the monomers.\cite{grest_review_brush} At time $t=0$ a constant
wall velocity of $v_{x}=1$ is switched on in one direction, and the
transient behavior of the brush is monitored. We took mean values
over 10 simulations starting from independent equilibrium configurations.
For each simulation, $3\cdot10^{5}$ to $6\cdot10^{5}$ steps with
a time step $dt=0.002$ were performed.

The rheological response of the polymer brush is analyzed for three
different cases: the usual Langevin thermostat with the same value
of $\gamma$ in all spatial directions, the Langevin thermostat with
zero friction constant in shear direction (denoted as $\gamma_{x}=0$),
and the DPD thermostat. It is known that the usual Langevin thermostat
($\gamma_{x}\neq0$) does not reproduce the hydrodynamic behavior
correctly because it does not conserve momentum\cite{dpd3,duenweg2}
and biases the flow profile in shear direction. A common workaround
to partially overcome these problems in non-equilibrium simulations
of simple (laminar) flows is to switch-off the Langevin thermostat
in the direction in which external non-conservative forces are applied.\cite{varsky_robbins,dpd3}
This corresponds to our second approach with $\gamma_{x}=0$. Alternatively,
Langevin thermostat is frequently switched-off in two Cartesian components,
being active only in the vorticity direction.\cite{priezjev1,thompson_robbins,grest_review_brush}

\begin{figure}
\begin{centering}
\includegraphics[bb=35bp 37bp 690bp 508bp,width=7cm]{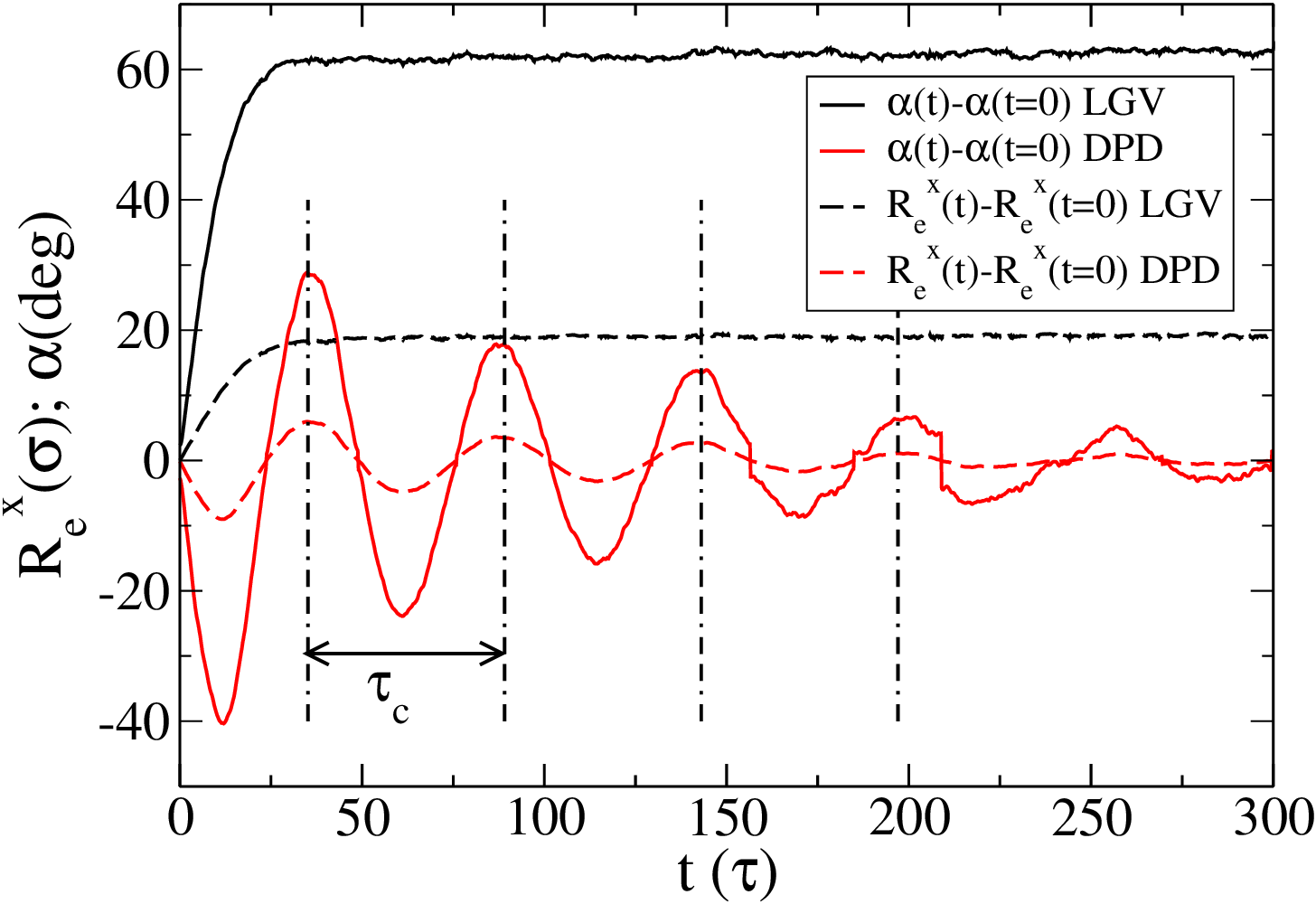} 
\par\end{centering}

\caption{(Color online) Transient evolution of the inclination angle $\alpha$
of the polymer brush (solid line) and the $\hat{x}$ component of
the end-to-end vector $R_{{\rm e}}^{x}$ (dashed line) for Langevin
and DPD thermostats. $\tau_{{\rm c}}$ is the characteristic response
time (see text and Fig.~\ref{fig:characteristic_time_sb}) \label{fig:single_brush:alpha_Re} }
\end{figure}

The thermostat's action can be rather understood as an implicit (ficticious)
solvent acting on the polymer beads. In our coarse-grained simulation
approach an effect of the implicit solvent on the dynamical properties
is undesirable. As we will show in the following, however, there are
cases where such effects cannot be avoided. In the case of the Langevin
thermostat, this solvent is at rest in the laboratory frame. When
the brush layer starts moving through the implicit solvent, the behavior
of Langevin and DPD thermostats differs drastically. A first evidence
of these differences is shown in Fig.~\ref{fig:single_brush:alpha_Re},
where Langevin and DPD thermostats are compared using a friction constant
of $\gamma=2$ in all spatial directions. The angle $\alpha$ between
the vector normal to the substrate and the average end-to-end vector,
${\bf R}_{{\rm e}}={\bf r}_{1}-{\bf r}_{N}$, with ${\bf r}_{1}$
and ${\bf r}_{N}$ denoting the position vectors of the grafted and
free end of a polymer chain, respectively, is shown as a function
of time. While for the DPD thermostat $\alpha$ exhibits a decaying
oscillatory behavior which ends in a steady state with the brush perpendicular
to the wall ($\alpha=0$), the Langevin thermostat shows an angle
which monotonously increases to a steady state value of $60^{o}$.
This can be understood in terms of the lack of Galilean invariance
of the Langevin thermostat.  The brush is dragged through a ficticious
solvent, which is always at rest in the laboratory frame. In case
of the DPD thermostat, the Galilean invariance, which follows from
momentum conservation, implies that the polymer brush at rest and
in steady state are equivalent. This case provides an example where
the particular thermostat implementation plays a crucial role for
the dynamics of the system and can even alter the results qualitatively.

\begin{figure}
\begin{centering}
\includegraphics[clip,width=7cm]{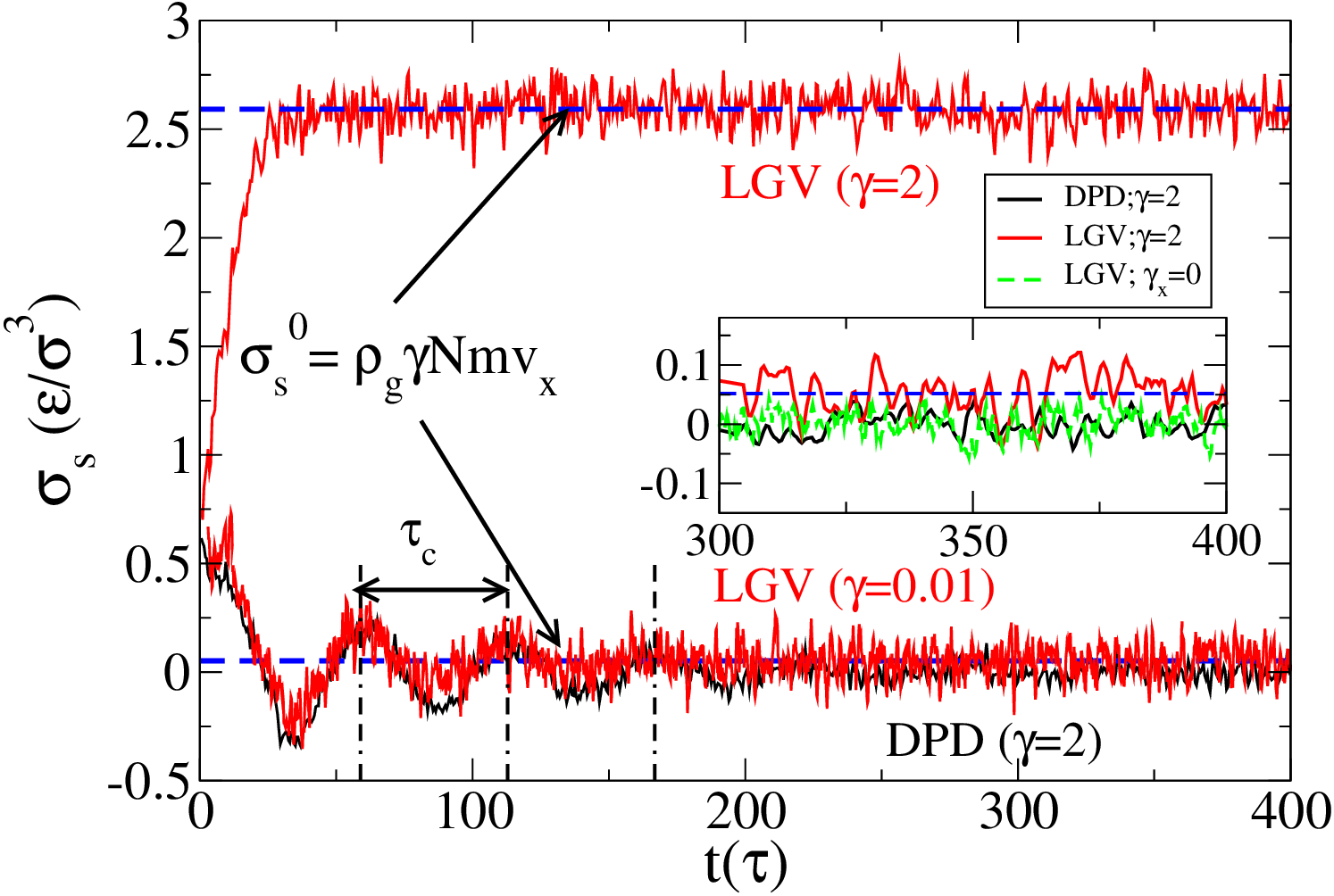} 
\par\end{centering}

\caption{(Color online) Comparison of shear stress for DPD and Langevin thermostats.
For the second case two different values of $\gamma$ were considered.
The inset shows the systematic difference for DPD and Langevin thermostat
with $\gamma=0.01$, close to steady state, and the approach to $\sigma_{{\rm s}}=0$
for the Langevin thermostat with $\gamma_{x}=0$ (dashed line). $\tau_{{\rm c}}$
is similar to Fig.~\ref{fig:single_brush:alpha_Re}. The mean stress
for Langevin is indicated with a horizontal dashed lines to improve
clarity.\label{fig:Fx_single_brush} }
\end{figure}

\begin{figure}
\begin{centering}
\includegraphics[clip,width=7cm]{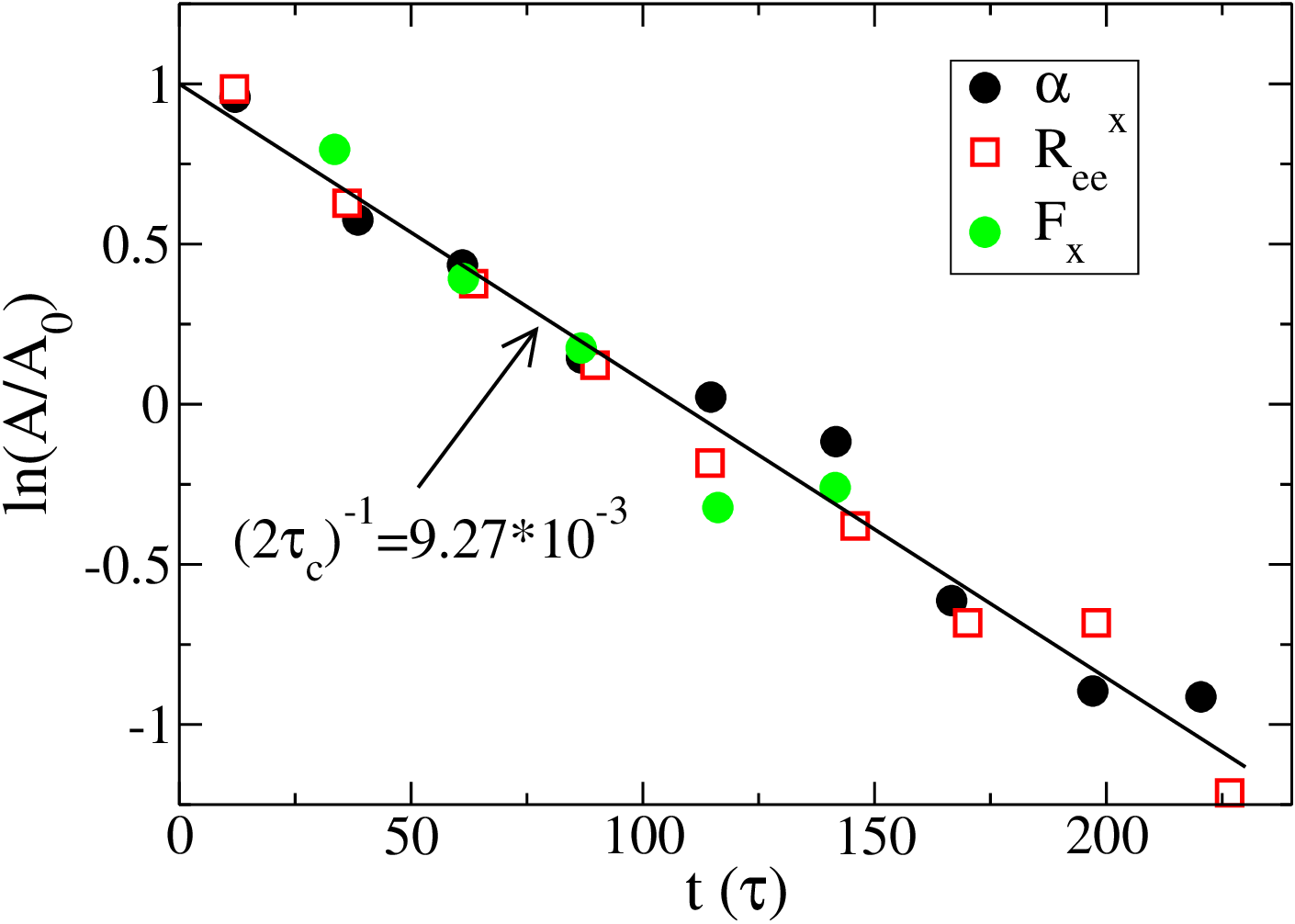} 
\par\end{centering}

\caption{(Color online) Maxima and minima of the oscillating curves (Figs.~\ref{fig:single_brush:alpha_Re}
and \ref{fig:Fx_single_brush}) for the single brush system. A characteristic
frequency of the system is found when using DPD or Langevin thermostats
with $\gamma=0.01$. \label{fig:characteristic_time_sb} }
\end{figure}

The oscillation frequency of $\alpha$ is a general property of the
brush layer from which we can extract a characteristic response time.
The component of the end-to-end vector in shear direction, $R_{{\rm e}}^{x}$
(dashed lines in Fig.~\ref{fig:single_brush:alpha_Re}), and the
mean shear stress (force per surface area) on the grafted heads of
the polymer brush (see Fig.~\ref{fig:Fx_single_brush}) present a
similar behavior. Defining for each observable $A\equiv A_{0}\exp(-t/\tau_{{\rm c}})$,
the decaying envelope of the oscillating curves, maxima and minima
can be brought onto the same curve (see Fig.~\ref{fig:characteristic_time_sb}),
yielding a characteristic time, $\tau_{{\rm c}}=53.94\tau$. The $\gamma_{x}=0$
case for the Langevin thermostat (not shown) presents the same behavior
as the DPD thermostat with the same characteristic time within the
error.

As observed for the mean thermostat forces (see Fig.~\ref{fig:Comparison-of-gammas})
in the good solvent case, similar forces for Langevin and DPD thermostats
are obtained for values of $\gamma$ which differ by two orders of
magnitude. We therefore performed the simulations for the Langevin
thermostat with $\gamma=0.01$ in all spacial directions, which corresponds
to $\gamma=2$ for the DPD case. Indeed, we observe the oscillations
also for the Langevin thermostat for a sufficiently small value of
$\gamma$ (see Fig.~\ref{fig:Fx_single_brush}). This shows that
the over-damping in the previous case ($\gamma=2$) is due to the
large, but not uncommon value of $\gamma$.

While the oscillations are reproduced for the smaller value of $\gamma$,
the steady state shear stress remains finite. It can easily be calculated
for both values of $\gamma$ via

\begin{equation}
\sigma_{{\rm s}}^{0}=\rho_{{\rm g}}\gamma Nmv_{x},\label{eq:sigma0}\end{equation}
 with $m\equiv1$ the monomer mass. As expected, the steady state
for $\gamma_{x}=0$ yields \textbf{$\sigma_{{\rm s}}=\sigma_{{\rm s}}^{0}=0$},
shown as a dashed line in the inset of Fig.~\ref{fig:Fx_single_brush}.

\begin{figure}
\begin{centering}
\includegraphics[clip,width=7cm,keepaspectratio]{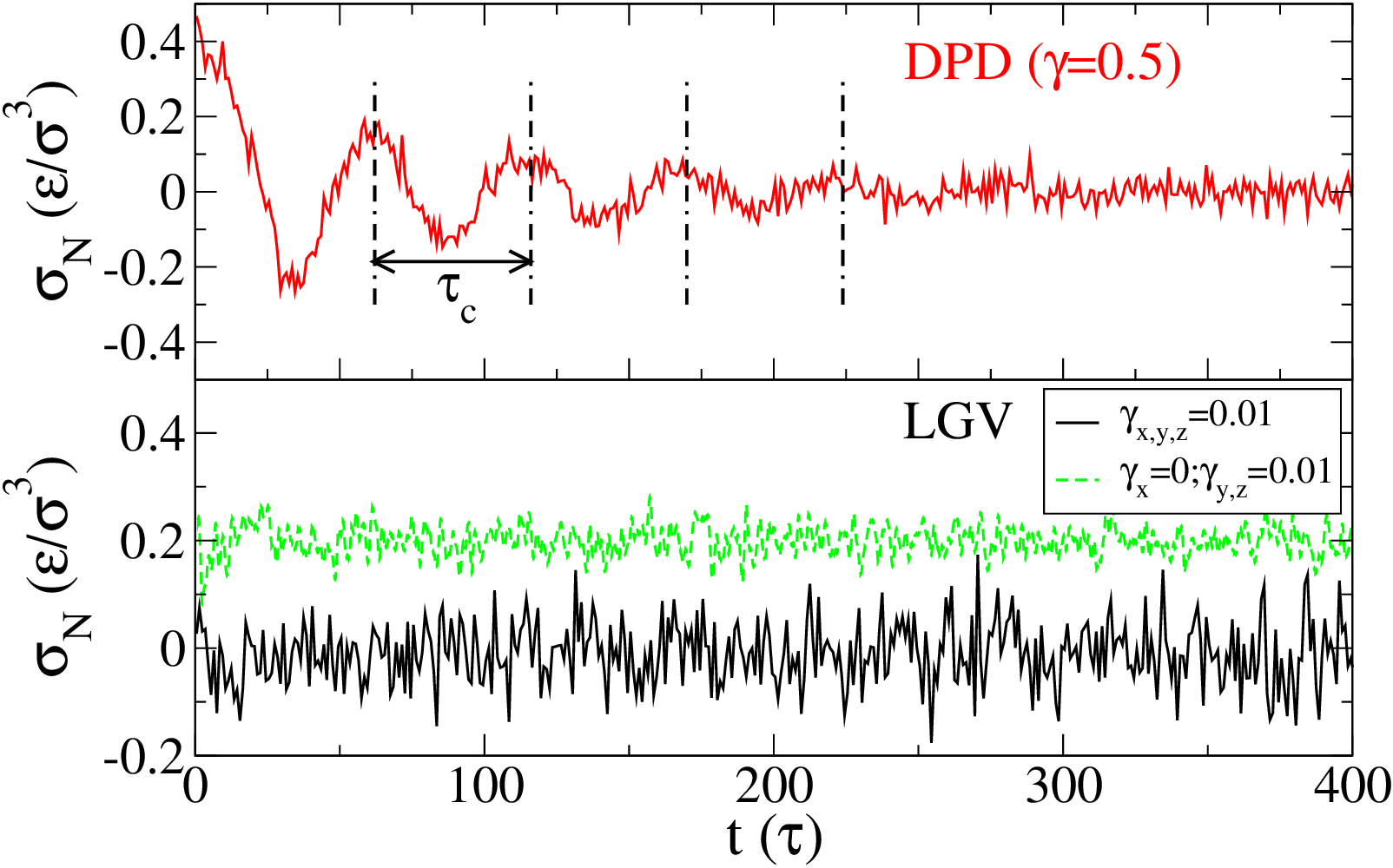} 
\par\end{centering}

\caption{(Color online) Upper panel: Normal stress as a function of time for
DPD. Lower panel: Langevin thermostat with $\gamma=0.01$ in all directions
and $\gamma_{x}=0$ along the shear direction. The curve corresponding
to the latter case was shifted by $0.2$ to improve clarity. \label{fig:Fz_single_brush}}
\end{figure}

Figure \ref{fig:Fz_single_brush} shows the transient behavior of
the normal pressure exerted by the brush layer. Again, oscillations
of the stress are observed for the DPD thermostat, while they are
not directly observed in either version of the Langevin thermostat
even for the smaller value of the friction coefficient. A Fourier
frequency analysis of the time sequence, however, exhibits a peak
at the same frequency for all cases.

At first one could argue that the Langevin thermostat over-damps the
oscillations because local momentum is only conserved in the shear
direction, and a non-equilibrium situation in which there is a strong
coupling between directions cannot be faithfully reproduced. This
coupling would be reinforced by the chain connectivity of the polymeric
system. We found, however, that for good solvent conditions and $\gamma=0.5$
the DPD thermostat is not able to maintain constant temperature with
the standard choice of weight functions. Taking, for example, weight
functions as $\omega^{{\rm R}}=\omega^{{\rm D}}=\Theta(r_{{\rm c}}-r)$
(see fourth column of table \ref{tab:Ntp}), the DPD thermostat maintains
the temperature for a wall velocity interval $v_{x}\in[0,1]$, but
also the normal pressure oscillations are suppressed.

\subsection{Two opposing brushes\label{sub:brush-brush}}

A system of two opposing brush layers was studied under constant shear.
It is similar to that already studied in previous works,\cite{Kreer2001,Kreer2003,Kreer2004,lai_binder_92,lai_brinder_93,binder_lai_94}
and originally attracted much interest because of its extraordinary
small lateral friction forces. The equilibrium density profiles are
depicted in Fig.~\ref{fig:Equilibrium-density-profiles}(b). We considered
brush layers of chains with $N=30$ beads at two different grafting
densities: $\rho_{{\rm g}}=1.2\rho_{{\rm g}}^{*}$ and $\rho_{{\rm g}}=4.9\rho_{{\rm g}}^{*}$,
where $\rho^{*}=1/\pi R_{{\rm g}}^{2}$ ($R_{{\rm g}}=3.02$ being
the radius of gyration of a single chain in solution) is the grafting
density characterizing the gradual crossover from the mushroom to
the brush regime. $\rho_{{\rm g}}=1.2\rho_{{\rm g}}^{*}$ is a system
within the crossover regime between mushroom and brush, whereas the
latter choice of $\rho_{{\rm g}}$ leads to a semi-dilute brush.\cite{Kreer2004}
For both considered grafting densities the opposing brushes interdigitate
because the distance between the opposing end-grafted beads is $D=17.5<2h_{0}$
($h_{0}$ denoting the unperturbed height of a single brush).

\begin{figure}
\begin{centering}
\includegraphics[width=7cm]{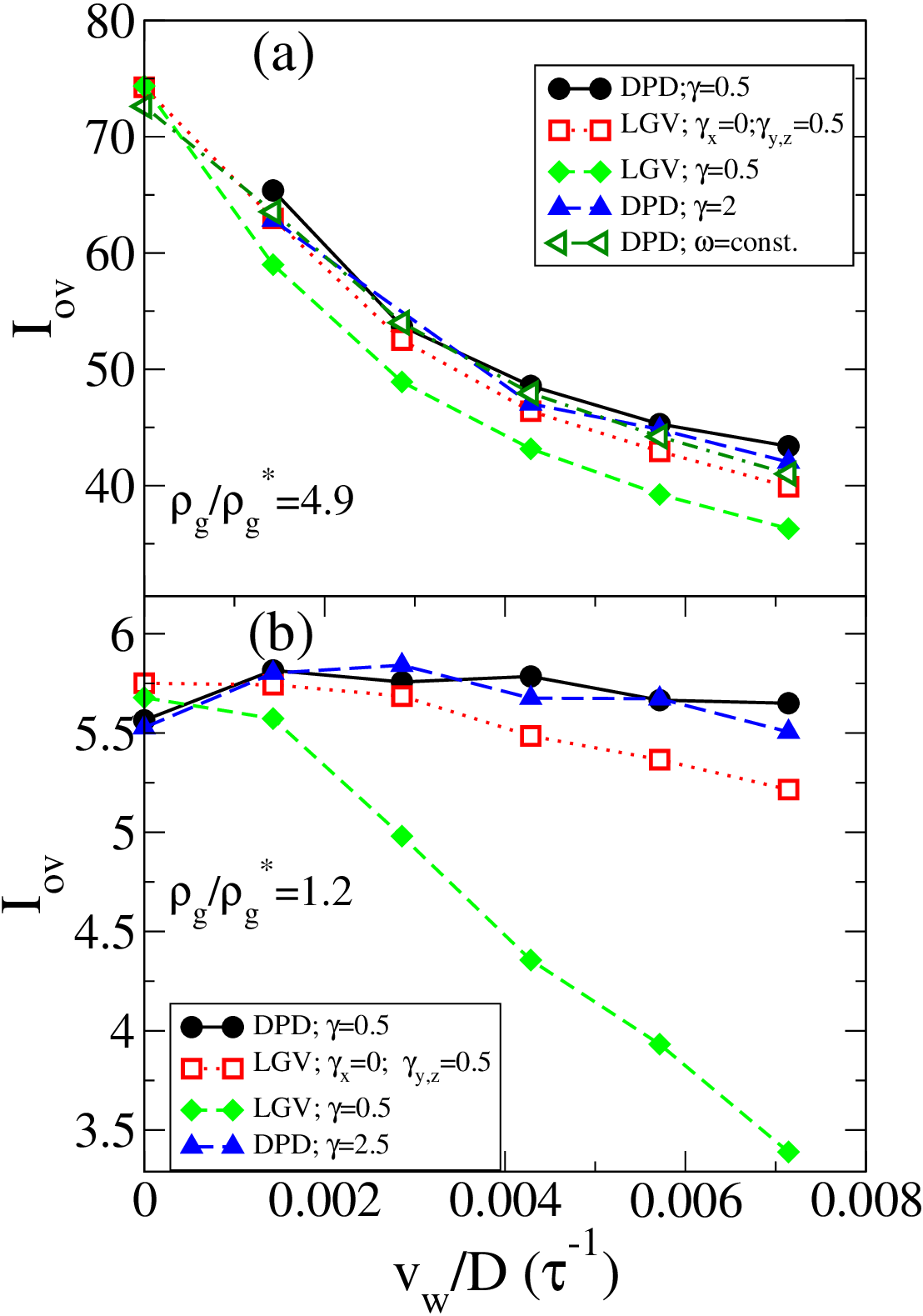} 
\par\end{centering}

\caption{(Color online) Overlap integral versus shear rate for the two studied
grafting densities. For DPD, we used the standard weight functions
{[}Eq.~(\ref{eq:usual-weight})] and the constant ones ($\omega={\rm const}$).}

\label{fig:Iov_vs_v} 
\end{figure}

Following previous works,\cite{Kreer2001,Kreer2003,claudio1} we quantify
the amount of interdigitation via the overlap integral, $I_{{\rm ov}}$,
defined as

\begin{equation}
I_{{\rm ov}}\equiv v_{{\rm mono}}A\int_{0}^{{\rm D}}\rho_{{\rm l}}(z)\times\rho_{{\rm u}}(z)dz\,,\label{eq:I_{ov}}\end{equation}
 where $\rho_{{\rm u}}$ and $\rho_{{\rm l}}$ are respectively the
number densities of upper and lower brush layers, $v_{{\rm mono}}{\bf =}0.52$
 is the volume of a monomer, and $A$ the surface area covered by
the grafted beads. $I_{{\rm ov}}$ follows from integrating the curves
depicted in the insets of Fig.~\ref{fig:Equilibrium-density-profiles}.
Figure \ref{fig:Iov_vs_v} shows this quantity as a function of the
shear rate, $v_{{\rm w}}/D$, for good solvent conditions. The interdigitation
is much higher for the larger grafting density, $\rho_{{\rm g}}=4.9\rho_{{\rm g}}^{*}$
{[}Fig.~\ref{fig:Iov_vs_v}(a)], as expected from the more important
stretching of the chains, and it becomes smaller with increasing shear
rate and the progressive tilting of the chains. For the Langevin thermostat
with $\gamma=0.5$ in all directions, this effect is more pronounced
because the biasing of the flow profile increases the tilting of the
brushes. A slight systematic difference can also be observed for the
DPD thermostat with $\gamma=0.5$ where the overlap is systematically
higher than in the other DPD cases. For $\gamma=0.5$ the temperature
is not properly conserved with the standard weight functions under
good solvent conditions and the brush is additionally stretched.

On the other hand, for $\rho_{{\rm g}}=1.2\rho_{{\rm g}}^{*}$, the
overlap is fairly constant for DPD and the Langevin thermostat with
$\gamma_{x}=0$ {[}Fig.~\ref{fig:Iov_vs_v}(b)] over the whole interval
of shear rates while for the standard Langevin thermostat ($\gamma_{x}\neq0$)
the strong monotonous decrease is again related to the bias in shear
direction.

\begin{figure}
\begin{centering}
\includegraphics[clip,width=7cm]{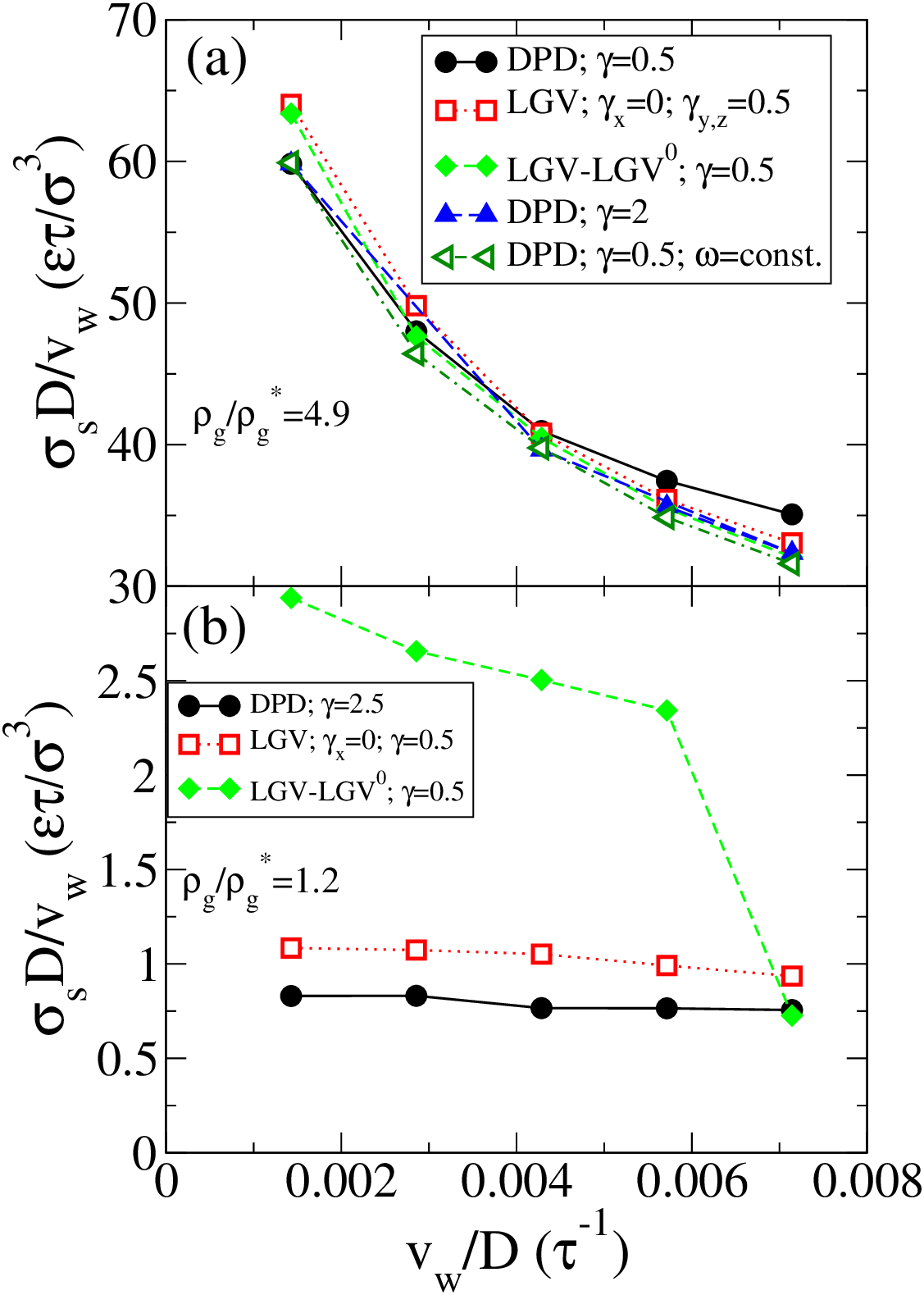} 
\par\end{centering}

\caption{(Color online) Comparison of {}``effective viscosity'' for two
opposing polymer brush layers under good solvent conditions for DPD
and Langevin thermostats, using Langevin damping with $\gamma_{x}\neq0$
and $\gamma_{x}=0$ . Different friction constants and two weight
functions for DPD are considered. High and low grafting densities
are presented in the upper and lower panel, respectively. \label{fig:shear_stress_bb} }
\end{figure}

Figure \ref{fig:shear_stress_bb}(a) shows the shear stress $\sigma_{{\rm s}}=\frac{\langle F_{x}\rangle}{A}$
($\langle F_{x}\rangle$ the mean force acting on the end-grafted
beads in shear direction) times the inverse shear rate as a function
of the constant relative wall velocity, $v_{w}$, for good solvent
conditions. As $\sigma_{{\rm s}}D/v_{w}$ reflects the {}``effective
viscosity'' of the polymeric system, the decrease of this quantity
indicates a non-linear behavior, known as shear-thinning. The overall
behavior of all cases is roughly the same. The Langevin thermostat
approximately reproduces the DPD result after subtracting the constant
shear stress given by the Langevin damping {[}Eq.~(\ref{eq:sigma0})]
from the measured value. The Langevin thermostat with $\gamma_{x}=0$
quantitatively agrees with the DPD case. The DPD thermostat was used
with the standard weight functions for two different values of the
friction constant: $\gamma=2$ and $\gamma=0.5$. The latter value
was also used in combination with constant weight functions. Only
for $\gamma=0.5$, the standard weight functions lead to some systematic
differences for the shear stress at large wall velocities. We found
out that this is due to the fact that under these conditions DPD fails
to maintain the temperature at the desired value ($T=1.68$).

Figure \ref{fig:shear_stress_bb}(b) shows the effective viscosity
for the smaller grafting density. The physical situation is now different
as compared to the previous case: the opposing brush layers have a
very small degree of interdigitation which is now independent of the
wall velocity {[}see also Fig. \ref{fig:Iov_vs_v}(b)].

The linear response is observed for DPD and Langevin thermostats with
$\gamma_{x}=0$. For the standard Langevin thermostat ($\gamma_{x}\neq0$),
however, the {}``effective viscosity'' decreases and drops drastically
for the largest wall velocity. This can be explained via the behavior
presented in section \ref{sub:Single-brush}: the tilting of the brush
reduces the interdigitation of the brush layers not only because of
the interaction among the brushes but also due to the strong interaction
with the ficticious solvent that the Langevin thermostat inevitably
implies. \textbf{}This is, however, an unphysical artifact in the
simulations studied here. As a consequence, the density profiles depend
on the parameters of the thermostat, which again is an unphysical
effect. \textbf{}Moreover, we emphasize that, even when the overall
behavior of DPD and Langevin thermostats with $\gamma_{x}=0$ is similar,
the absolute value of the shear stress is different. %

\begin{figure}
\begin{centering}
\includegraphics[clip,width=7cm]{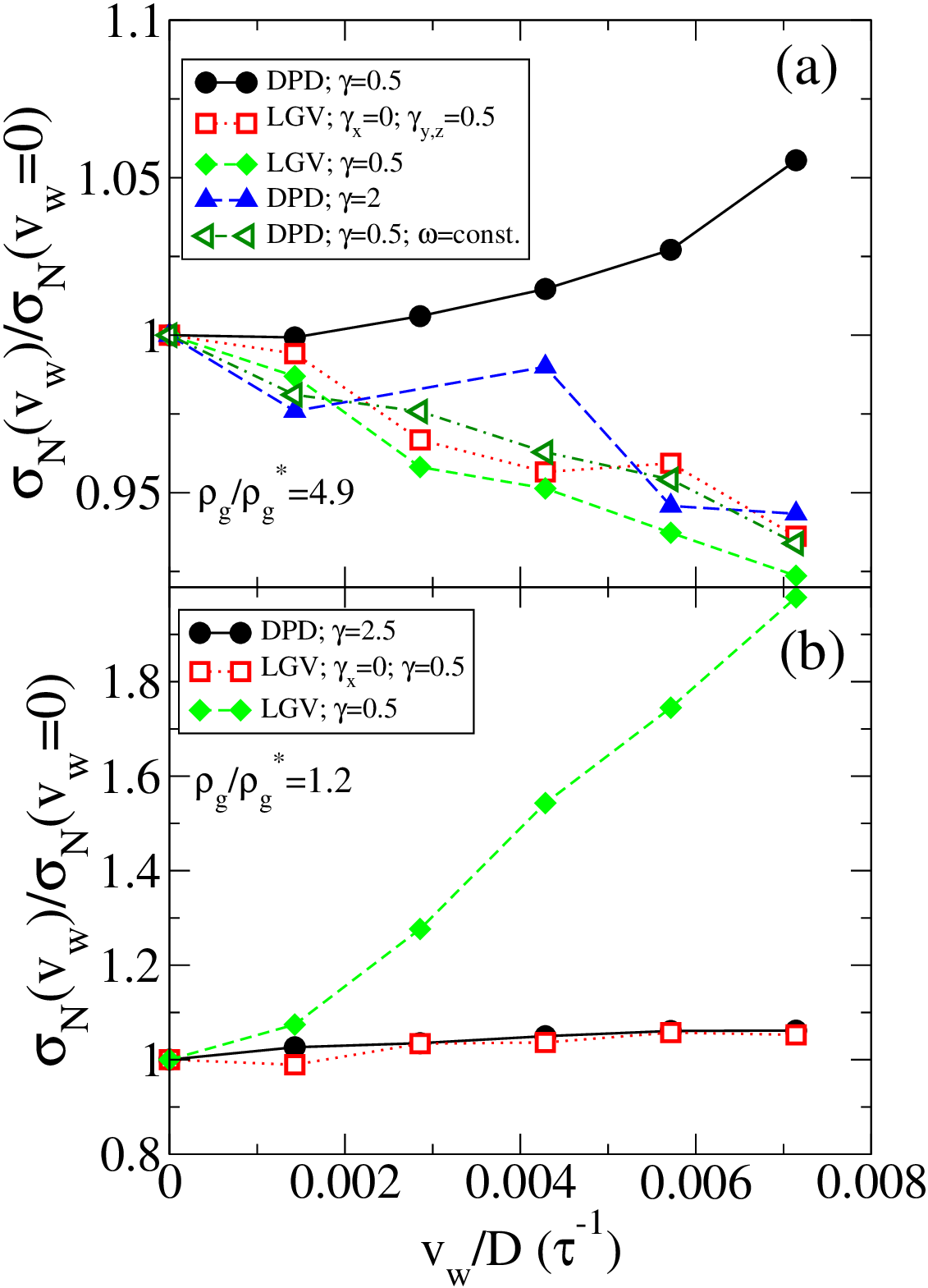} 
\par\end{centering}

\caption{(Color online) Comparison of normal pressure for two opposing polymer
brush layers under good solvent conditions. DPD with different values
of friction constants, $\gamma$, and weight functions are considered
and compared to the standard Langevin thermostat ($\gamma$ equal
in all spatial directions) and the case $\gamma_{x}=0$. The upper
panel shows a grafting density with strong interdigitation between
the brushes, and the lower one presents a case with small brush-brush
interdigitation.\label{fig:normal_stress_bb} }
\end{figure}

Figure \ref{fig:normal_stress_bb} shows the normal stress as a function
of shear rate for the two studied grafting densities. Figure \ref{fig:normal_stress_bb}(a)
considers the higher grafting density, for which the brushes are strongly
interdigitated and slightly compressed (under good solvent conditions
the mean force between the layers is repulsive). A decreasing normal
stress as a function of shear velocity is found, except for the case
of DPD with $\gamma=0.5$ and the standard weight functions. As mentioned
above, for this case DPD does not properly conserve temperature and
a slight heat up of the system is observed, which in turn produces
a further increase in the steady state stretching of the brush with
a concomitant increase in the normal repulsion of the brush layers.
For all the other cases, the decrease of normal pressure upon increasing
velocity is produced by the progressive tilting of the chains and
the decrease of interdigitation, already observed in the behavior
of the overlap integral {[}Fig. \ref{fig:Iov_vs_v}(a)].

Figure \ref{fig:normal_stress_bb}(b) shows the normal stress for
the smaller grafting density. Here, the brush is so dilute that the
interaction between the brush layers is almost negligible. In this
case, there is a mean attraction between the layers, due to the wall
interaction with each bead {[}see Eq.~(\ref{eq:v_wall})]. For DPD
($\gamma=2.5$) and the Langevin thermostat with $\gamma_{x}=0$,
the structure of the brush is quite similar resulting in a similar
behavior of the normal force. The small interdigitation leads to a
very small change of the inclination angle giving rise to a very weak
dependence of the normal stress on the wall velocity. This behavior
agrees with the approximately constant behavior of the overlap integral
shown in Fig.~\ref{fig:Iov_vs_v}(b) and the linear response observed
in the effective viscosity. A different behavior is found for the
Langevin case with $\gamma=0.5$ in all directions. Under these conditions,
the unphysical enhancement of the chain inclination leads to a much
larger interaction between wall and monomers, which is mainly attractive
for typical bead positions.

Normal and shear stresses were also investigated under poor solvent
conditions. We used the Langevin thermostat with $\gamma=0.5$ perpendicular
to the shear direction and $\gamma_{x}=0$. For the DPD thermostat
with $\gamma=0.5$, the temperature is conserved under poor solvent
conditions unlike in the good solvent case. As already shown in table
\ref{tab:Ntp}, the larger cut-off radius for poor solvent conditions
improves the DPD efficiency and keeps $T$ constant even for the standard
weight functions.

\begin{figure}
\begin{centering}
\includegraphics[clip,width=7cm]{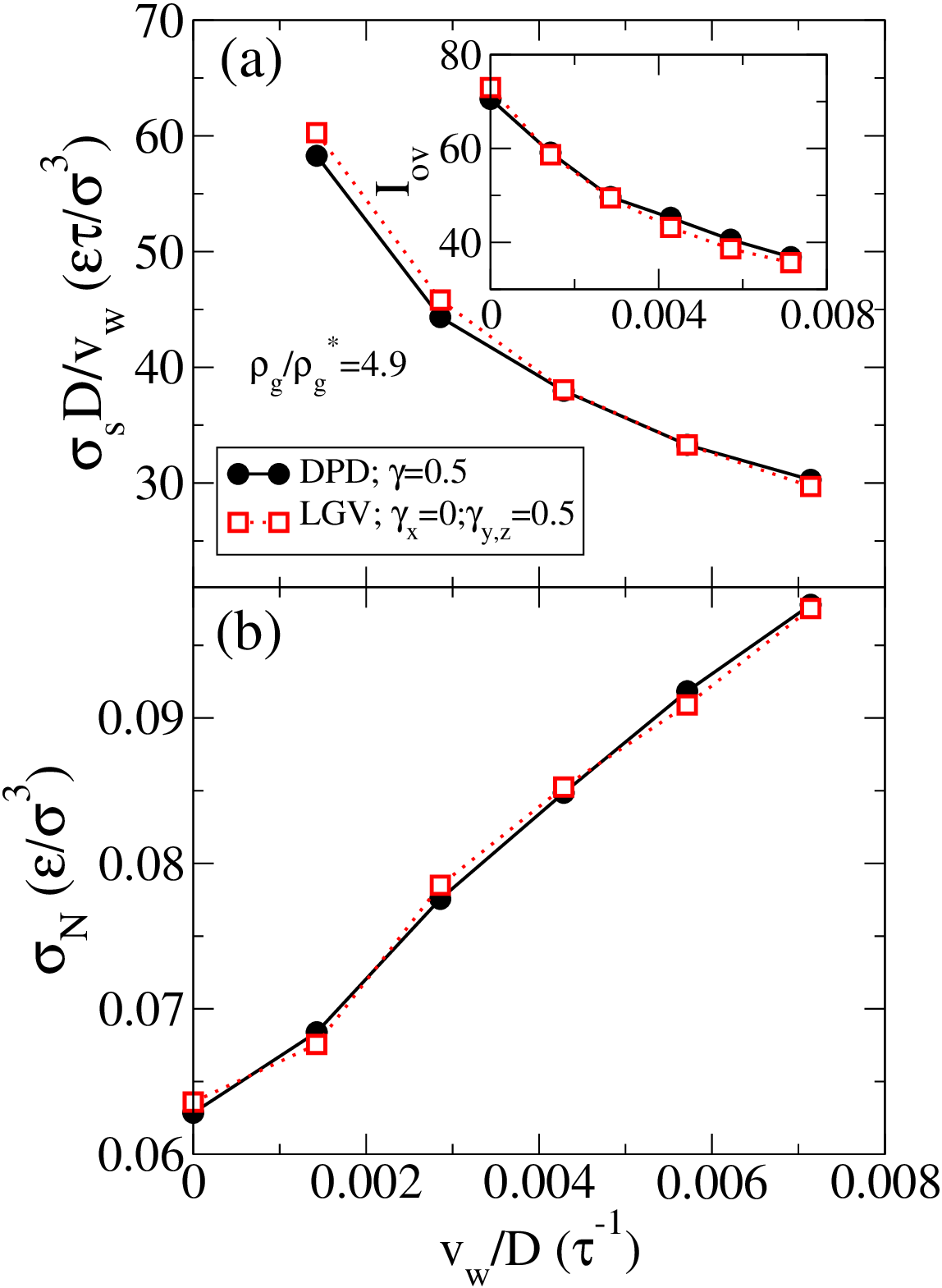} 
\par\end{centering}

\caption{(Color online) Comparison of shear and normal stress for DPD and
Langevin thermostat with $\gamma_{x}=0$ (poor solvent conditions).
Both thermostats give a similar result. The inset shows the overlap
integral $I_{{\rm ov}}$ as defined in Eq.~(\ref{eq:I_{ov}}).\label{fig:shear_stress_ps_bb} }
\end{figure}

Figure \ref{fig:shear_stress_ps_bb}(a) shows a very similar behavior
of both thermostats concerning the effective viscosity, $I_{ov}$,
and the normal stress. Within the considered regime of shear rates,
both thermostat implementations lead to equivalent results. As compared
with the good solvent case, the normal stress at the wall is negative,
i.e. there is a mean attraction between upper and lower brush layers.
This is due to the mean attraction among beads for this model.

If this equivalence between DPD and LGV with $\gamma_{x}=0$ holds
also for stronger out-of-equilibrium situations, i.e at higher shear velocities,
remains to be investigated thoroughly.

\subsection{The role of weight functions in the DPD thermostat\label{sub:role_weight_functions}}

In this section we consider, in more detail, the ability of the DPD
thermostat to conserve temperature in non-equilibrium simulations
for different weight functions. The system under study is a polymeric
liquid, formed by 10-bead chains, confined by two polymer brushes
of identical chains.\cite{claudio1} The brush and melt density profiles
across the perpendicular directions correspond to that of Fig.~\ref{fig:Equilibrium-density-profiles}(c).
We imposed either a Poiseuille flow by means of a constant external
volume force or a linear Couette flow by moving the walls at constant
relative velocity.

\begin{figure}
\begin{centering}
\includegraphics[clip,width=7cm]{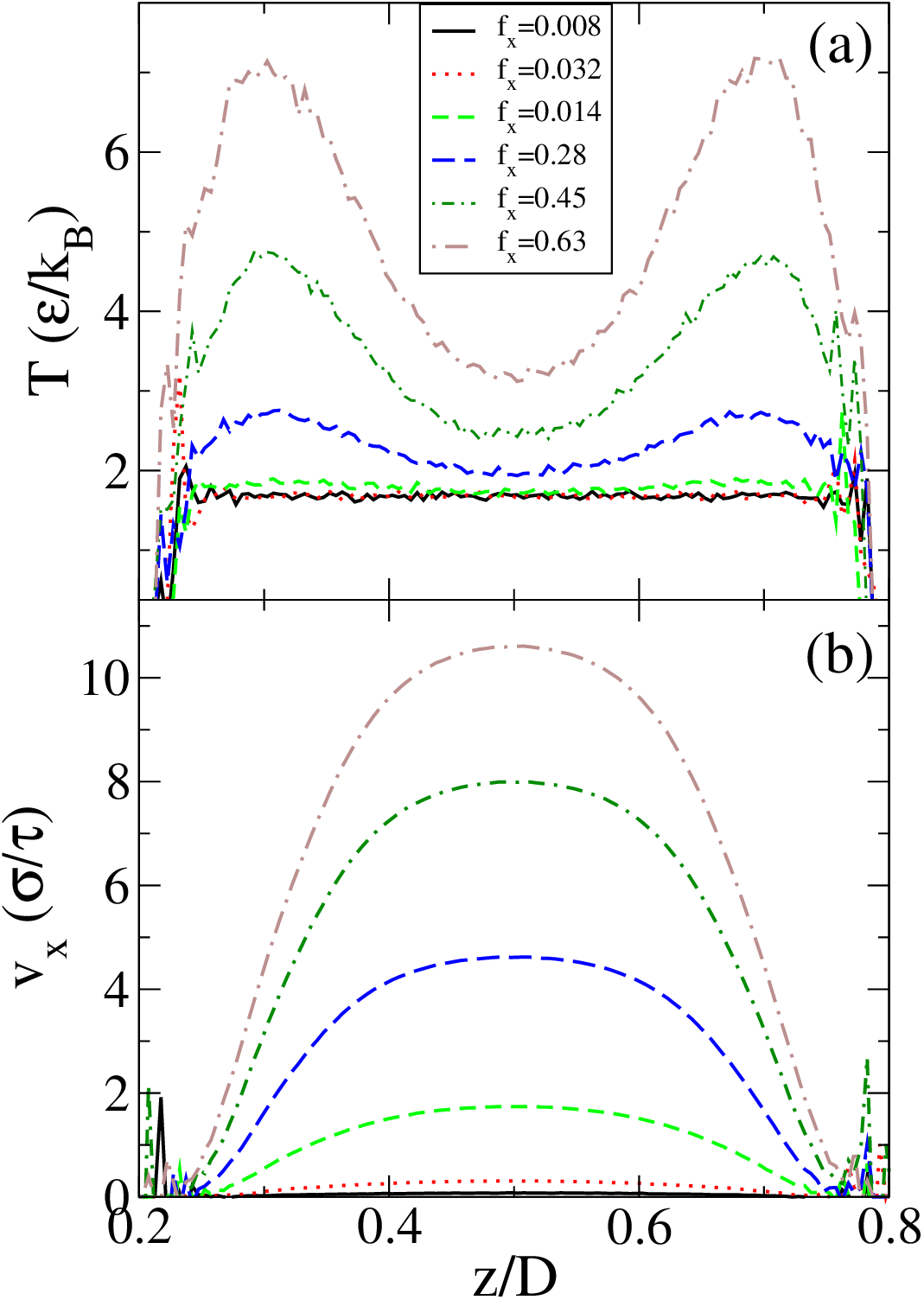} 
\par\end{centering}

\caption{(Color online) Poiseuille flow for different external forces, $f_{x}$.
Upper panel: Temperature, as obtained from the mean square velocity
perpendicular to the shear direction. Lower panel: Velocity profile
across the polymeric liquid. Only for the two smallest external forces
DPD with the standard weight functions {[}Eq.~(\ref{eq:usual-weight})]
maintains the temperature at the desired value. The simulations were
performed under poor solvent conditions.\label{fig:dpd_weight_poiss}}
\end{figure}

In Fig.~\ref{fig:dpd_weight_poiss}, the violation of temperature
conservation is shown as a function of external force, $f_{x}$, for
Poiseuille flow using the standard DPD weight functions. The temperature,
liquid number density, and brush grafting density were respectively
set to $T=1.68$, $\rho_{{\rm m}}=0.61$ and $\rho_{{\rm g}}=5.5\rho_{{\rm g}}^{*}$
($=0.77\sigma^{-2}$), with $\rho^{*}=1/\pi R_{{\rm g}}^{2}$ and
$R_{g}=1.50$. The typical Poiseuille velocity profile across the
gap is shown in Fig.~\ref{fig:dpd_weight_poiss}(b), while the temperature
profile -- measured by the mean square velocity in the direction perpendicular
to the flow -- is presented in Fig.~\ref{fig:dpd_weight_poiss}(a).
Temperature is conserved only for the two smallest forces. In the
remaining cases, the temperature increases in the region of large
velocity gradients. These cases show examples in which the DPD thermostat
fails to maintain the desired temperature even under poor solvent
conditions.

\begin{figure}
\begin{centering}
\begin{tabular}{c}
\includegraphics[clip,width=7cm]{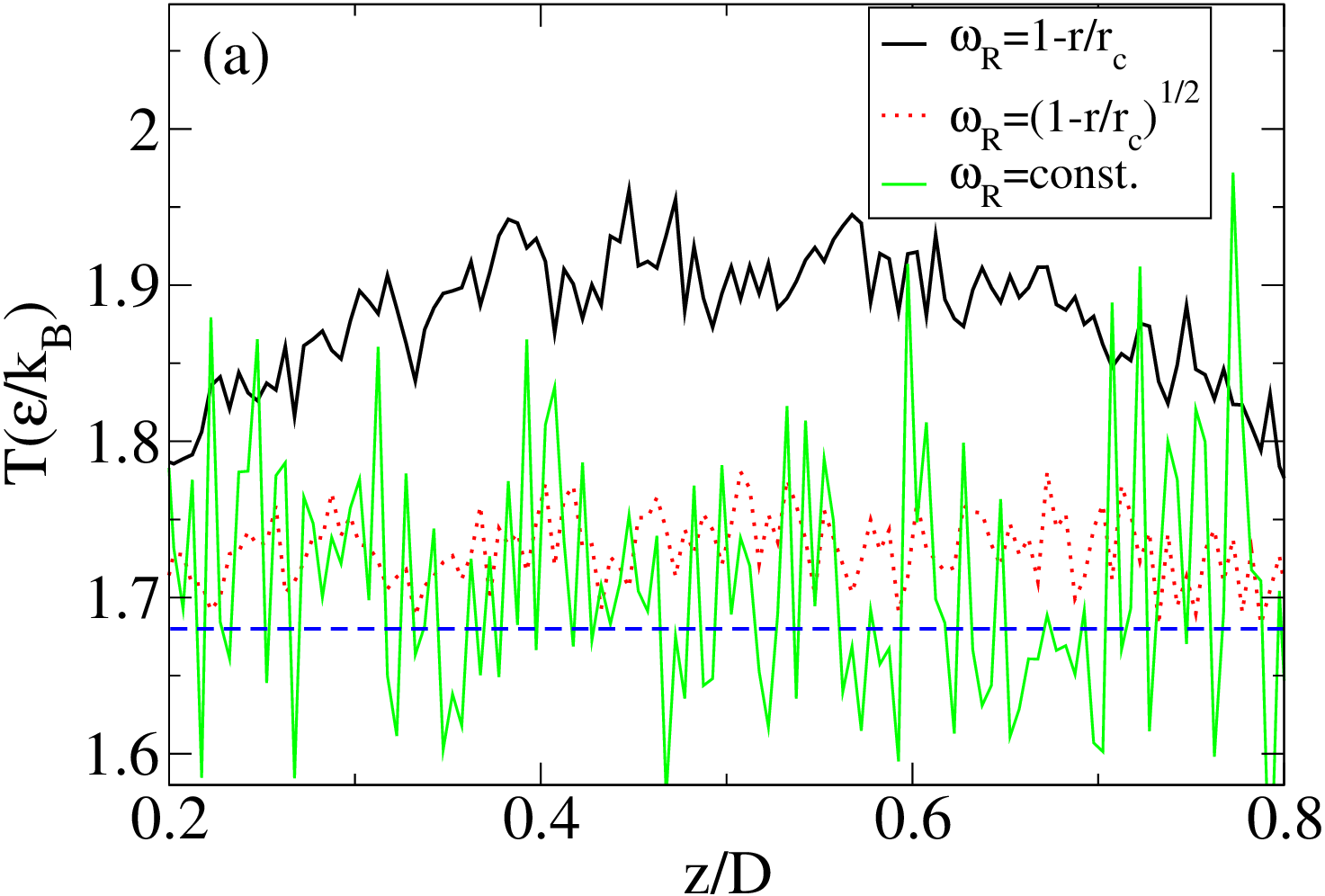}\tabularnewline
\includegraphics[clip,width=7cm]{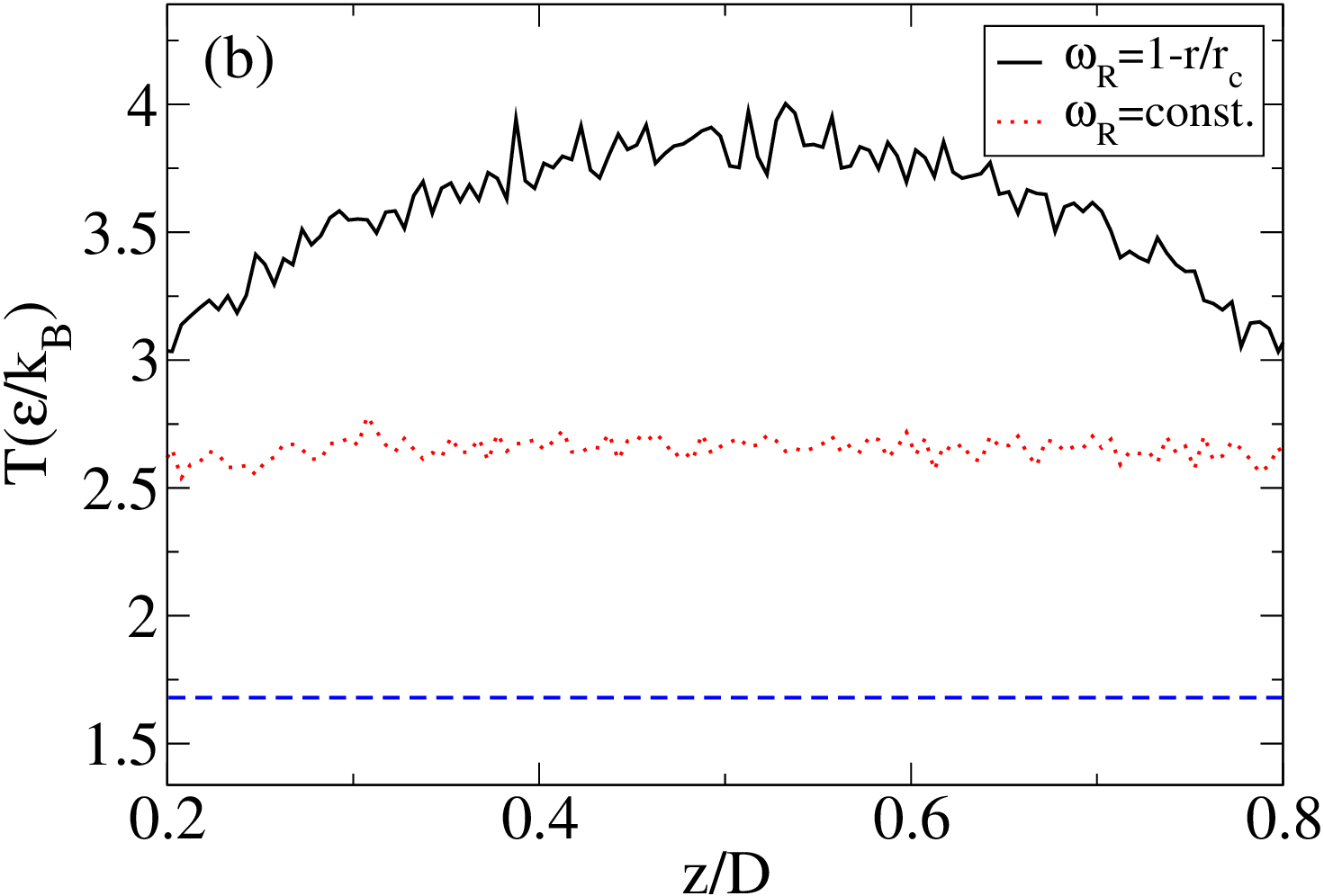}\tabularnewline
\end{tabular}
\par\end{centering}

\caption{(Color online) Comparison of different DPD weight functions for a
Couette flow (poor solvent conditions). In the upper panel, $v_{w}=3$
is considered: while the standard choice of DPD weight functions is
not able to conserve temperature, a choice of $\omega_{R}=(1-\frac{r}{r_{{\rm c}}})^{1/2}$
improves temperature conservation, and $\omega_{R}=\Theta(r_{{\rm c}}-r)$
conserves temperature at the chosen value (indicated by the dashed
line). At $v_{w}=8$ (lower panel ), temperature is not conserved
at the desired value, and the behavior across the film is very different
for the considered weight functions (see text). \label{fig:temp_prof_dpd_weights}}
\end{figure}

We also studied the temperature conservation for Couette flows using
the weight functions considered in table \ref{tab:Ntp}. In Fig.~\ref{fig:temp_prof_dpd_weights}(a)
the temperature profiles for a shear velocity of $v_{w}=3$ are shown.

The standard weight functions clearly fail to keep temperature constant
and lead to quadratic temperature profiles with a maximum in the middle
of the gap. In contrast to the Poiseuille flow this is not related
to the velocity gradient, which is constant across the gap for Couette
flows. We attribute the resulting temperature profiles to the density
distribution of monomers, which is enhanced close to the brush coated
walls giving rise to a local improvement of the efficiency of the
thermostat in this region. Another choice of weight functions {[}$\omega^{{\rm R}}=\sqrt{\omega^{{\rm D}}}=(1-r/r_{{\rm c}})^{1/2}$]
gives a better result although it also fails to conserve temperature.
The constant-weight functions ($\omega^{{\rm R}}=\sqrt{\omega^{{\rm D}}}=1$)
are more efficient and conserve temperature.

Figure \ref{fig:temp_prof_dpd_weights}(b) shows the temperature profile
for a shear velocity of $v_{w}=8$. Under this condition, the thermostat
is not able to conserve temperature because for reasonable values
of the friction constant energy cannot be dissipated as fast as it
is plugged into the system. The solid line, corresponding to the standard
DPD weight functions, shows a similar behavior as in the previous
case: temperature is conserved more efficiently in the regions of
higher density. For constant weight functions, temperature is fairly
constant all across the film but differs from the desired value (indicated
by a dashed line).

>From these examples of strong out-of-equilibrium simulations, we conclude
that the choice of weight functions can make an important difference.
For example, the standard weight functions might not be able to conserve
temperature even for a physically meaningful case. Constant weight
functions seem to be a good alternative and are even more efficient
computationally, as already noted in a previous study.\cite{dpd3}
As a general guideline, it should be checked that the physical conditions
fulfill the relation $\dot{\gamma}=N_{\rm TP}\gamma_{\rm DPD}/m$, where $\dot{\gamma}$
is the shear stress imposed to the system. In any case, for short
range potentials (as Lennard-Jones and good solvent conditions) or strong out-of-equilibrium
simulations, the temperature profile across the sample, as shown in
Figure \ref{fig:temp_prof_dpd_weights} can give insight on the ability
of the thermostat for keeping the temperature constant.

\section{Conclusions\label{sec:Conclusions}}

In this work we tested and compared commonly used implementations
of Langevin and DPD thermostats for different polymeric systems. Equilibrium,
transient and steady state conditions were considered for the study
of various reference systems, such as polymer brush bearing surfaces
or brush-melt interfaces. We utilized a well studied coarse-grained
bead-spring polymer model. By varying the cut-off in the interaction
potential we mimicked good and poor solvent conditions. We quantified
the relative strength of the thermostats in a wide range of friction
constants, $\gamma$, and found that the strength of the Langevin
thermostat is much larger than DPD with standard weight functions
for similar values of $\gamma$. The simulation of the transient state
of a polymer brush layer driven to constant velocity from rest illustrates
the known weaknesses of the Langevin thermostat -- lack of momentum
conservation, screening of hydrodynamic interactions, and violation
of Galilean invariance -- and how these are avoided by the DPD thermostat
which conserves local momentum. When applied in shear direction, the
Langevin thermostat biases the velocity profile. The common workaround
of switching-off the Langevin thermostat in the non-equilibrium direction
was analyzed for different systems. We found that, in most cases,
the latter behaves similar to the DPD scheme but care must be exerted
when the system is strongly driven out of equilibrium.

We furthermore quantified the differences between various forms of
weight functions of the DPD thermostat, which can be chosen freely,
provided that the weights for random and dissipative forces obey relation
(\ref{eq:dpd_weights_rel}). It is important to note that most of
the previous works using DPD utilized a standard form {[}Eq.~(\ref{eq:usual-weight})],
which was originally intended to be used in conjunction with {}``soft''
potentials. When the DPD thermostat is applied to typical coarse-grained
potentials, the {}``hard'' nature of the conservative potentials
prevents one from using a very large time step, and therefore not
much is gained from {}``smooth'' thermostat forces. Moreover, we
found that the typical weights can be regarded as adequate for equilibrium
and slightly out-of-equilibrium conditions but they fail to conserve
temperature for medium and strong driving forces. We tested this for
both, Poiseuille and Couette flows, of a polymer melt confined between
two polymer brush layers. We found quantitatively that taking constant
weight functions is both computationally faster and yields a thermostat
that is more suitable for strong out-of-equilibrium situations in
which a large amount of heat per unit time is produced.

It is important to note that none of the methods discussed in the
present paper is suitable to fully account for hydrodynamic effects
at arbitrary densities. For instance, the coupling between the monomer
density distribution and the external flow profile is not described.
To achieve this, other methods, e.g., the self-consistent solution
of the Brinkman equation\cite{brinkman} have to be applied. In a
future study we plan to investigate the systems considered here using
explicit solvent molecules. This should put us into the position to
understand the importance of the effects delineated above.

Finally, our results clearly show that great care is needed in non-equilibrium
MD simulations of soft matter systems in order to ensure that the
simulations are free of artifacts due to an inappropriate choice of
the thermostat.

\begin{acknowledgments}
It is a great pleasure to thank Hendrik Meyer for useful discussions
and driving our attention on the role of the weight functions in the
DPD thermostat. Jörg Baschnagel and Joachim Wittmer are also gratefully
acknowledged. Financial support by the DFG within the priority program
{}``Micro- und Nanofluidik'' Mu 1674/3-1, the Sonderforschungsbereich
625/A3, the ESF-program STIPOMAT, and the DAAD/SECYT are gratefully
acknowledged. Computing time was provided by the NIC, Jülich, Germany.
C.P. also thanks ANPCYT (PME 2003, PICT 2005) for financial support.\\

\end{acknowledgments}

 \bibliographystyle{apsrev}


\begin{thebibliography}{40}
\bibitem[{bd_()}]{bd_sd}
\bibinfo{note}{The regime in which the accelerations are kept and are
  integrated explicitly is in general called Stochastic Dynamics.
  Alternatively, if friction and random forces are big enough, the
  instantaneous values of the accelerations can be set to zero. This is
  typically referred to as Brownian Dynamics.}
\end{thebibliography}

\end{document}